\documentclass[oneside,reqno]{amsart}
\usepackage{amsmath,amsthm,amssymb,amsfonts}
\usepackage{mathrsfs}      
\begin{document}
\pagenumbering{arabic} \numberwithin{equation}{section}
\theoremstyle{remark}\newtheorem{Example}{{\bf Example}}[section]
\newcommand{\overcirc}[1]{\overset{\circ}{#1}}
\newcommand{\smashovercirc}[1]{\smash[t]{\overset{\circ}{#1}}}
\newcommand{\dummy}{\protect \rule{0.0em}{1.0ex}}
\newcommand{\Dummy}{\protect \rule{0.0em}{1.6ex}}
\newcommand{\preone}{\,\Dummy^{(1)}}
\newcommand{\pretwo}{\,\Dummy^{(2)}}
\newcommand{\prethree}{\,\Dummy^{(3)}}
\newcommand{\bbar}{\protect \raisebox{-.2ex}[1ex][0ex]{\hspace*{0.1ex}\protect \rule{.08ex}{1.3ex}\hspace*{.25ex}\protect \rule{.08ex}{1.3ex}\hspace*{.3ex}}}
\newcommand{\highstar}{\protect \raisebox{0.5ex}{*}}
\newcommand{\tzero}{\overcirc{t}_0}
\title[Cumulative gravitational lensing]{Cumulative gravitational lensing in Newtonian perturbations of Friedman-Robertson-Walker cosmologies}
\author[C.-M. Claudel]{Clarissa-Marie Claudel$\Dummy^1$}
\address{Mathematical Institute, University of Ume{\aa}, S-901 87 Ume{\aa}, Sweden.}
\begin{abstract}
It is a common assumption amongst astronomers that, in the
determination of the distances of remote sources from their
apparent brightness, the cumulative gravitational lensing due to
the matter in all the galaxies is the same, on average, as if the
matter were uniformly distributed throughout the cosmos. The
validity of this assumption is considered here by way of general
Newtonian perturbations of Friedman-Robertson-Walker (FRW)
cosmologies. The analysis is carried out in synchronous gauge,
with particular attention to an additional gauge condition that
must be imposed.  The mean correction to the apparent
magnitude-redshift relation is obtained for an arbitrary mean
density perturbation. In the case of a zero mean density
perturbation, when the intergalactic matter has a dust equation of
state, then there is
 indeed a zero mean
first order correction to the apparent magnitude-redshift relation
for all redshifts. Point particle and Swiss cheese models are
considered as particular cases.
\end{abstract}
\maketitle
\protect\addtocounter{footnote}{1}
\protect\footnotetext{\noindent e-mail: {\tt
clarissa@abel.math.umu.se}}
\section{Introduction}
The purpose of this paper is threefold: first to give a general
analysis of synchronous gauge Newtonian perturbations of
Friedman-Robertson-Walker
(FRW) cosmologies; second to obtain a condition
which
identifies a suitable homogenised FRW cosmology for comparison
purposes, and third to obtain the relevant formula for the
correction to the apparent magnitude-redshift relation for weak
gravitational lensing.

The classic paper of Lifschitz \& Khalatnikov (1963) generated much interest
in the study of perturbations of FRW cosmologies.  Among the influential
papers that followed were those of Hawking (1966) and Sachs \& Wolfe (1967).
A review that covers both classical and quantum aspects is
Mukhanov et al.\ (1992).  One of the main topics in the literature is the
growth of small perturbations in the early universe and the formation of the
galaxies.  The present work, by contrast, is concerned with the description of
the universe as it is today, with the galaxies providing an essentially time
independent perturbation of an FRW background.

Although the use of a gauge invariant approach is frequently advocated,
since it eliminates any need to specify and interpret gauge conditions, it is
nonetheless more appropriate here to employ a synchronous gauge.  This gauge
is known to involve a certain ambiguity, as was pointed out by Mukhanov et
al.\ (1992), and has given rise to difficulties of interpretation
(Press \& Vishniac 1980).  It will therefore be necessary in the present work
to identify an appropriate supplementary gauge condition which admits a
natural interpretation. This will be done in \S\ref{gauge}.

Gravitational lensing has been considered by many authors and
with different techniques (see e.g. Schneider {\it et al.} 1992;
Bertotti 1966; Dyer \& Roeder 1972,1973, 1974; Kantowski 1998).
One such technique has been the use of stochastic numerical integration
of the lensing equations (Holz \& Wald 1998; Dyer \& Oattes 1988).
The effects of weak gravitational lensing on the cosmic microwave
background radiation have been considered by Seljak (1996) and by
Sachs \& Wolfe (1967). The optical properties of the Swiss cheese
model of Einstein \&  Strauss (1945,1946) have been considered by
Dyer \& Roeder  (1974).  Both the Swiss cheese model and the point particle
model  of Newman \& McVittie (1982) will be considered here as
applications of a general theory.

Newtonian perturbations of FRW cosmologies have previously been
considered by McVittie (1931), Harrison (1967) and Newman \& McVittie (1982).
A central difficulty is that if one neglects terms involving the spatial
curvature constant $k$, the total gravitational potential for a uniform
distribution of  particles will be everywhere infinite.  What does not seem to
be  widely appreciated is that this problem disappears when proper account is
taken of either a positive or negative $k$. This will be discussed in
{\S}\ref{ex}$\,b$.

Much of the analysis will be carried out in the linear
approximation in the gravitational coupling constant $\kappa =
8\pi G/c^2$. Consequently only weak gravitational lensing will be
described. In particular, shear is neglected because it has a
second order effect on luminosity. Caustics are also neglected.
These may both be significant restrictions. There are now
well-known examples of astronomical images that show the effects
of caustics, although there are as yet no firm estimates of the
proportion of images for which caustics are involved.  It has been
claimed by Ellis {\it et al.} (1998) that light rays followed back
from our present location in time and space meet a galaxy within a
redshift $0< z<5$.  If this is correct then strong gravitational
lensing is likely to be significant for images with $z\geq 5$.

The question as to whether, in general, weak gravitational lensing
gives the same apparent magnitude-redshift relation on average as
a best-fit FRW model was considered by Weinberg (1976) who gave two
supporting arguments.  The first was based on an analysis of double
image lensing but was only valid for $q \ll 1$, i.e. for $\Omega \ll 1$.
The second, subsequently generalised by Peacock (1986), was essentially
based on photon conservation (Schneider {\it et al.} 1992, pp.99--100),
and was valid for all $\Omega$.  Although widely accepted in the
literature, these arguments have been criticised by Ellis {\it et al.}
(1988), the first on the grounds that generic lensing gives rise
to three images, not two, and the second at least in part because
it effectively assumes the result to be proved.  The criticism of
Weinberg's first argument is presumably based on the odd number theorem
of Burke (1981) and McKenzie (1985), according to which a transparent
gravitational lens can give rise only to an odd number of images.
However this theorem is flawed since, as was pointed
out by Gottlieb (1994), it attempts to use $3$-dimensional topology
in a $4$-dimensional setting.  Nonetheless the criticism of Weinberg's
second argument appears to stand.  The present paper therefore gives a
new supporting argument that, in the weak field limit, and when
averaged over large angular scales, the apparent magnitude-redshift
relation does indeed agree with that of a best-fit FRW model.

\section{General perturbations of FRW cosmologies}\label{genpert}
The FRW metric, in standard form, is
\begin{equation}
\overcirc{\bf g} = -dt^2 + S^2 (t)\overcirc{\bf h} \label{gFRW}
\end{equation}
where $\overcirc{\bf h}$ is a $t$-independent 3-metric given by
\begin{equation}
\overcirc{\bf h} = \left( 1+
\frac{kr^2}{4}\right)^{-2}\sum_{i=1}^3 (dx^i)\Dummy^2 \
.\label{hFRW}
\end{equation}
Here and henceforth, indices $i,j\ldots$ run from $1$ to $3$,
whilst indices $a,b \ldots$ will run from $0$ to $3$. For the most
part it will be possible to describe the level surfaces of $t$
with respect to an arbitrary coordinate system ${\bf x}= \{
x^1,x^2,x^3\}$. The space-time may then be conveniently described
with respect to the coordinate system $\{ x^a :a=0,\ldots ,3\}$
with $x^0=t$.

Quantities associated with $\overcirc{\bf g}$ will carry a
superscript $\overcirc{\dummy}$\ . Covariant differentiation with
respect to $\overcirc{\bf g}$ will be denoted by $\overcirc{;}$,
e.g.\ $\overcirc{g}_{ab\smashovercirc{;}c}=0$. The energy tensor
of $\overcirc{\bf g}$ is
\begin{equation}
\overcirc{T}\Dummy^{ab} = (\overcirc{\rho} + \overcirc{p})u^a u^b
+ \overcirc{p}\overcirc{g}\dummy^{ab} \label{TFRW}
\end{equation}
where $\overcirc{g}\dummy^{ab}$ is the inverse of
$\overcirc{g}_{ab}$, $u^a :=
-\overcirc{g}\dummy^{ab}t_{\smashovercirc{;}b}$ is the unit
future-directed normal to the level surfaces of $t$,
\begin{align}
\kappa \overcirc{\rho}(t) & = \frac{3k}{S^2 (t)} +
\frac{\dot{S}\Dummy^2 (t)}{S^2 (t)} - \Lambda  \label{rhoFRW} \\
\kappa \overcirc{p}(t) &= - \frac{k}{S^2 (t)} -
\frac{\ddot{S}(t)}{S(t)} + \Lambda \label{pFRW}
\end{align}
give the density $\overcirc{\rho}(t)$ and pressure
$\overcirc{p}(t)$ respectively, and $\Lambda$ is the cosmological
constant, which is not assumed to vanish.  A superscript dot
denotes differentiation with respect to $t$.

The Hubble and deceleration parameters for $\overcirc{\bf g}$ are
defined by
\begin{align}
\overcirc{H}(t)& := \frac{\dot{S}(t)}{S(t)}  \label{Hdef} \\
\overcirc{q}(t) &
:= -\frac{\ddot{S}(t)S(t)}{\dot{S}\Dummy^{2}(t)}  \label{qdef}
\end{align}
respectively whilst the dimensionless density parameter is defined
by
\begin{equation} \overcirc{\Omega}(t) := \frac{\kappa
\overcirc{\rho}(t)}{3\overcirc{H}\Dummy^2 (t)}\ .
\label{denpardef}
\end{equation}
Note that by (\ref{rhoFRW}) and (\ref{pFRW}) one has
\begin{equation}
\frac{d}{dt}\overcirc{H}(t)  = - \overcirc{H}\Dummy^2 (t)(1 +
\overcirc{q}(t)) = \frac{k}{S^2 (t)} - \frac{1}{2}\kappa
(\overcirc{\rho} + \overcirc{p}) \ . \label{Hdot0}
\end{equation}

Covariant differentiation with respect to $\overcirc{\bf h}$ in
the level surfaces of $t$ will be denoted by a subscript
$\overcirc{\protect\bbar}$, e.g.
$\overcirc{h}_{ij\smashovercirc{\protect\bbar}k}=0$. Quantities
associated with the 3-metric $\overcirc{{\bf h}}$, apart from
$\overcirc{{\bf h}}$ itself will, in addition to the superscript
$\overcirc{\dummy}$\ , carry a prefix $\prethree$.  Thus the
Laplacian associated with $\overcirc{\bf h}$ will be denoted by
$\prethree\overcirc{\Delta}$.   The Riemann and Ricci tensors and
Ricci scalar of $\overcirc{\bf h}$ are
\begin{align}
\prethree \overcirc{R}\Dummy^i\Dummy_{jkl} & =
2k\delta^i\dummy_{[k}\overcirc{h}\dummy_{l]j} \label{3Riemh0} \\
\prethree\overcirc{R}_{ij} & = 2k\overcirc{h}_{ij}
\label{3Riccih0} \\ \prethree\overcirc{R} & = 6k \label{3Rh0}
\end{align}
respectively.

The calculations of the remainder of this section pertain to a
perturbation of the FRW metric (\ref{gFRW}) of the particular
form
\begin{equation}
{\bf g} = -dt^2 + S^2 (t){\bf h} \label{gdef} \end{equation}
where
\begin{equation}
{\bf h} = \overcirc{\bf h} + \delta {\bf h} \label{hdef}
\end{equation}
is a $3$-metric intrinsic to the level surfaces of $t$, i.e.\ transverse
to $\overcirc{{\bf u}}$, and where $\delta {\bf h}$ may depend upon
all four coordinates.

The scale factor $S(t)$, the cosmological constant $\Lambda$ and
the spatial curvature constant $k$ will all be taken to be the
same as for the unperturbed metric $\overcirc{\bf g}$. Note that
since $\overcirc{\bf h}$ is independent of $t$ one has $\partial_t
h_{ij} = \partial_t \delta h_{ij}$.  Covariant differentiation
with respect to ${\bf h}$ in the level surfaces of $t$ will be
denoted by $\protect\bbar$, e.g.\ $h_{ij\protect\bbar k}= 0$.
Quantities associated with ${\bf h}$, other then ${\bf h}$ itself,
will carry a prefix $\prethree$. Thus the Laplacian associated
with ${\bf h}$ will be denoted by $\prethree\Delta$ and the Ricci
tensor and Ricci scalar of ${\bf h}$ will be denoted by $\prethree
R_{ij}$ and $\prethree R$ respectively. Indices $i,j\ldots $ on
quantities associated with ${\bf h}$ will be raised and lowered by
${\bf h}^{-1}$ and ${\bf h}$ respectively; indices $a,b\ldots$ on
quantities associated with ${\bf g}$ will be raised and lowered by
${\bf g}^{-1}$ and ${\bf g}$ respectively. Differences between
quantities associated with ${\bf g}$ and ${\overcirc{\bf g}}$, or
${\bf h}$ and ${\overcirc{\bf h}}$, will be denoted by $\delta$,
e.g.\ $\delta \prethree R_{ij} := \prethree R_{ij} - \prethree
\overcirc{R}_{ij}$.

The vector field $u^a = -t^{;a}$ is irrotational and geodesic. The
integral curves of ${\bf u}$ are to be regarded as the world lines
of a preferred family of observers, described henceforth as
comoving. In particular the galaxies are presumed to be comoving.
The coordinate $t$ thus carries physical significance.

The second fundamental form of the level surfaces of $t$ is given
by
\begin{equation}
\chi_{ab} := u_{a;b} \ . \label{sffdef}
\end{equation}
Clearly $\chi_{ab}$ is transverse to $u^a$ and so defines a
$3$-tensor intrinsic to the level surfaces of $t$:
\begin{equation}
\chi_{ij} = \Gamma^t_{ij} = \frac{1}{2}\partial_{t}(S^2(t)h_{ij})
= \dot{S}(t)S(t)h_{ij} + \frac{1}{2}S^2(t)\partial_t\delta h_{ij}
\ . \label{sff}
\end{equation}
The Hubble and deceleration parameters of $u^a$ with respect to $g_{ab}$
are defined by
\begin{align}
H & := u^a\dummy_{;a} = \frac{1}{3}\chi^a\dummy_{a} =
\overcirc{H}(t) + \frac{1}{6}h^{ij}\partial_t \delta h_{ij}
\label{exp} \\
q & := \partial_t \left( \frac{1}{H} \right) -1  \ . \label{qdefgen}
\end{align}
Note that $H$ and $q$ are functions of all four coordinates.  By
(\ref{sff}) the Gauss equation for $\prethree R^i\Dummy_{jkl}$
leads to
\begin{align}
\prethree R &=6k -S^2(t)\kappa (\overcirc{\rho} - \overcirc{p}) +
2S^2(t)(\kappa \delta T_{tt} -
\frac{\dot{S}(t)}{S(t)}h^{ij}\partial_t \delta h_{ij} )\nonumber
\\ & \quad + \frac{1}{2}h^{i[m}h^{j]n}(\partial_t\delta
h_{ij})(\partial_{t}\delta h_{mn}) \ . \label{3R}
\end{align}
From (\ref{sff}), (\ref{exp}), (\ref{qdefgen}) and (\ref{3R}) one has
\begin{align}
\delta \chi_{ij} &= \frac{1}{2} \partial_{t}(S^2(t) \delta h_{ij})
\label{deltasff} \\
\delta H &= \frac{1}{6}
h^{ij}\partial_{t}\delta h_{ij} \label{deltaexp} \\
\delta q &= - \frac{1}{\overcirc{H}}\partial_t \left( \frac{\delta H}{H}
\right) - (1+\overcirc{q})\frac{\delta H}{H}  \label{deltaq} \\
\delta\prethree R & = 2S^2(t)(\kappa \delta T_{tt} -
\frac{\dot{S}(t)}{S(t)}h^{ij}\partial_t \delta h_{ij} ) +
\frac{1}{2}h^{i[m}h^{j]n}(\partial_t  \delta  h_{ij})(\partial_t\delta h_{mn})
\ .  \label{delta3R}  \end{align}
The components of the Ricci tensor of $g_{ab}$ are given by
\begin{align}
R_{tt} &= -3\frac{\ddot{S}(t)}{S(t)} - \frac{1}{2S^2(t)}\partial_t
(S^2(t)h^{ij}\partial_t\delta h_{ij}) -
\frac{1}{4}h^{ik}h^{jl}(\partial_t\delta h_{ij})(\partial_t\delta
h_{kl}) \label{Rttg}\\ R_{it}=R_{ti}&=h^{jk}(\partial_t\delta
h_{k[i})\dummy_{\protect\bbar j]} \label{Ritg}\\ R_{ij}& =
(2k+2\dot{S}\Dummy^2(t)+S(t)\ddot{S}(t) +
\frac{1}{2}\dot{S}(t)S(t)h^{km}\partial_t \delta h_{km})h_{ij}
\nonumber \\ & \quad + \delta\prethree R_{ij} - 2k\delta h_{ij}
 + \frac{1}{2S(t)}\partial_t(S^3(t)\partial_t\delta
h_{ij} ) \nonumber \\ & \quad+
\frac{1}{4}S^2(t)h^{km}(\partial_t\delta h_{km})(\partial_t\delta
h_{ij}) - \frac{1}{2}S^2(t)h^{km}(\partial\delta
h_{ik})(\partial_t\delta h_{jm}) \ , \label{Rijg}
\end{align}
the last of these having been obtained by means of
(\ref{3Riccih0}).  One also has
\begin{align}
R&=6\bigl( \frac{k}{S^2(t)} + \frac{\dot{S}\Dummy^2(t)}{S^2(t)} +
\frac{\ddot{S}(t)}{S(t)}\bigr) + \frac{1}{S^2(t)}\delta \prethree
R \nonumber \\ & \quad + \frac{1}{S^4(t)}\partial_t
(S^4(t)h^{ij}\partial_t\delta h_{ij})
+\frac{1}{2}h^{i(j}h^{k)m}(\partial_t \delta h_{km})(\partial_t
\delta h_{ij}) \label{Rg}
\end{align}
by means of (\ref{3Riccih0}) and
\begin{equation}
\delta\prethree R = h^{ij}\prethree R_{ij} -
\overcirc{h}\Dummy^{ij}\prethree \overcirc{R}_{ij} = h^{ij}(\delta
\prethree R_{ij} - 2k\delta h_{ij}) \ . \label{delta3R1}
\end{equation}
Hence the energy tensor ${\bf T}$ of ${\bf g}$, as defined by
Einstein's equations
\begin{equation} \kappa T_{ab} = R_{ab} -\frac{1}{2}Rg_{ab} +\Lambda
g_{ab} \ ,\end{equation} has components
\begin{align}
\kappa T_{tt} &= \kappa \overcirc{\rho} +
\frac{1}{2S^2(t)}\delta\prethree R +
\frac{\dot{S}(t)}{S(t)}h^{ij}\partial_t\delta h_{ij} +
\frac{1}{4}(h^{ij}\partial_t\delta h_{ij})^2 \label{Tgtt}\\ \kappa
T_{it} = \kappa T_{ti}& = (h^{jk}\partial_t\delta
h_{k[i})_{\protect\bbar j]} \label{Tgit} \\ \kappa T_{ij} &=  \{
S^2(t)\kappa\overcirc{p} -
\frac{1}{2S(t)}\partial_t(S^3(t)h^{km}\partial_t\delta
h_{km})\nonumber \\ & \quad \quad \quad
-\frac{1}{4}S^2(t)h^{p(q}h^{k)m}(\partial_t\delta
h_{km})(\partial_t\delta h_{pq})\} h_{ij} \nonumber\\ & \quad
+\delta \prethree R_{ij} -2k\delta h_{ij} -
\frac{1}{2}h_{ij}\delta\prethree R + \frac{1}{2S(t)}\partial_t
(S^3(t)\partial_t\delta h_{ij}) \nonumber \\ & \quad
+\frac{1}{4}S^2(t)h^{km}\{ (\partial_t \delta h_{km})(\partial_t
\delta h_{ij} ) - 2(\partial_t \delta h_{ik})(\partial_t\delta
h_{jm})\} \label{Tgij}
\end{align}

The Weyl tensor $C^a\Dummy_{bcd}$ of $g_{ab}$ may be decomposed in
the standard form
\begin{equation} C_{abcd} = 8u_{[a}E_{b][d}u_{c]} +
2g_{a[c}E_{d]b} -2g_{b[c}E_{d]a}
-2u^{e}\eta_{eabf}u_{[c}B_{d]}\Dummy^f -
2u^e\eta_{ecdf}u_{[a}B_{b]}\Dummy^f
\end{equation}
where $\eta_{abcd}$ is the alternating tensor,
\begin{align}
E_{ab} & := C_{acbd}u^cu^d \label{Eabdef} \\ B_{ab} &:= \highstar
C_{acbd}u^cu^d \label{Habdef}
\end{align}
are the electric and magnetic parts of $C^a\Dummy_{bcd}$, and
\begin{equation}
\highstar C_{abcd} := \frac{1}{2}\eta_{abef}C^{ef}\Dummy_{cd} =
\frac{1}{2}\eta_{cdef} C^{ef}\Dummy_{ab}
\end{equation} is the dual of $C^a\Dummy_{bcd}$.

In order to compute $E_{ab}$ one may use the identity
\begin{equation} 2u^bu_{a;[bc]} =
R_{dabc}u^bu^d \end{equation} to obtain
\begin{equation}
E_{ij} = \frac{1}{2}R_{ij} - \frac{1}{2}S^2(t)(R_{tt} +
\frac{1}{3}R)h_{ij} - \frac{D}{\partial t}\chi_{ij} -
\frac{1}{S^2(t)}\chi_{ik}\chi^k\Dummy_{j} \end{equation} which by
(\ref{sff}), (\ref{Rttg}), (\ref{Rijg}) and (\ref{Rg}) gives
\begin{align}
E_{ij} &= -\frac{1}{4}S(t)\{\partial_t (S(t)\partial_t \delta
h_{ij}) - \frac{1}{3}h_{ij}h^{km}\partial_t (S(t)\partial_t \delta
h_{km})\} \nonumber \\
& \quad + \frac{1}{2}(\delta \prethree R_{ij} - 2k\delta h_{ij} -
 \frac{1}{3}h_{ij}\delta \prethree R ) \nonumber \\
& \quad + \frac{1}{8} S^2(t)\{ h^{km}(\partial_t \delta
h_{km})(\delta_t \delta h_{ij})
- \frac{1}{3}h^{np}h^{km}(\partial_t \delta
 h_{km})(\partial_t\delta h_{np}) h_{ij} \} \ .
\label{Eijg} \end{align}

To compute $B_{ab}$ one may proceed directly from (\ref{Habdef})
to obtain
\begin{equation}
B_{ab} = -u^e\eta_e\Dummy^{cd}\Dummy_a ( \chi_{bd;c} -
\frac{1}{2}g_{cb}R_{fd}u^f )\end{equation} which by the symmetry
of $B_{ab}$ gives
\begin{equation}
B_{ab} = -u^e\eta_e\Dummy^{cd}\Dummy_{(a}\chi_{b)d;c} \ .
\end{equation}
By means of (\ref{sff}) one thus has
\begin{equation}
B_{ij} = -\frac{1}{4}S(t)\{ \prethree
\eta^{lm}\Dummy_i(\partial_t \delta h_{jm})_{\protect\bbar l} + \prethree
\eta^{lm}\Dummy_j(\partial_t\delta h_{im})_{\protect\bbar l} \} \label{Hijg}
\end{equation} where
\begin{equation} \prethree \eta_{ijk} =
\frac{1}{S^3(t)}u^a\eta_{aijk} \label{alth} \end{equation} is the
alternating tensor for $h_{ij}$.

The calculations so far have all been exact.  However,
in order to facilitate further progress, the perturbation
$\delta{\bf h}$ will be treated as a power series in $\kappa$ that
vanishes to zeroth order, and the perturbed metric ${\bf g}$ will be
studied only to first order in $\kappa$.  Note that the spatial
curvature constant $k$ will not be treated as a power series in
$\kappa$ so, for example, terms with the coefficient $k\kappa$
will not be disregarded in first order approximations.  This
contrasts, in particular, with the work of Holz \& Wald (1998).

\section{Redshift and emission time} \label{red}
Suppose a comoving observer with world line
$ \{ {\bf x} = {\bf x}_1\}$ in the perturbed space-time makes an
observation at time $t=t_1$ of a comoving source with redshift $z_1$.
The image seen by the observer is formed by a congruence of null geodesics
centred on a null geodesic $\gamma$ connecting a point $(t_0,{\bf x}_0)$ on
the source world line $\{ {\bf x}={\bf x}_0\}$ to the observation point
$(t_1,{\bf x}_1)$ on the observer's world line $ \{ {\bf x} = {\bf x}_1\}$.
For comparison, consider an observer in the unperturbed   space-time who, at
the same observation point $(t_1,{\bf x}_1)$, makes an   observation of a
comoving source with the same redshift $z_1$.   In this case let
$\overcirc{\gamma}$ be the null geodesic,
$ \{ {\bf x}= \overcirc{\bf x}\dummy_0\} $ the source world line and
$\tzero$ the time of emission.  For definiteness let $\overcirc{\gamma}$  be
chosen such that its spatial direction at $(t_1,{\bf x}_1)$  coincides with
that of $\gamma$.

The immediate problem is to compute the perturbation $\delta t_0=
t_0 - \overcirc{t}\dummy_0$ in the time of emission.
For $\overcirc{\bf g}$ the emission time $\tzero$ is
determined implicitly by
\begin{equation} 1 + z_1 = \frac{S(t_1)}{S(\tzero)}
\label{red0}
\end{equation}
which for nearby sources gives
\begin{equation}
t_1 -\tzero = \frac{z_1}{\overcirc{H}(t_1)} + O(z_1^2) \quad
\mbox{for $z_1 \ll 1$} \ . \label{red0approx}
\end{equation}
For ${\bf g}$ one may determine $t_0$ as follows.  Since $\gamma$
is a null geodesic one has
\begin{align}
S^2(t)h_{ij}\frac{dx^i}{dt}\frac{dx^j}{dt} & = 1 \label{null1} \\
\frac{d^2x^a}{d\lambda^2} + \Gamma^a_{bc}
\frac{dx^b}{d\lambda}\frac{dx^c}{d\lambda} & = 0  \label{geodesic}
\end{align}
where $\lambda$ is an affine parameter along $\gamma$ with value
$\lambda_0$ at $(t_0,{\bf x}_0)$ and value $\lambda_1$ at
$(t_1,{\bf x}_1)$.
By means of (\ref{null1}) the $t$ component of (\ref{geodesic}) gives
\begin{equation}
\frac{d}{dt}\ln \left( \frac{dt}{d\lambda} \right) +
\frac{\dot{S}(t)}{S(t)} + \kappa \nu_{\gamma}(t) = 0 \label{d2t}
\end{equation}
where
\begin{equation}
\kappa \nu_{\gamma}(t) := \frac{1}{2}S^2(t)(\partial_t \delta
h_{ij})\frac{dx^i}{dt}\frac{dx^j}{dt}  \label{nudef}
\end{equation}
is defined along $\gamma$.  The redshift as observed at time $t$
is given by the standard formula
\begin{equation}
 1+z(t) =
 \frac{\left( \left. \frac{dt}{d\lambda}\right|_{t_0} \right) }{\left( \left.
\frac{dt}{d\lambda}\right|_{t}\right)} \label{redshift}
\end{equation}
which combines with (\ref{d2t}) to give
\begin{equation}
 \frac{d}{dt}\ln (1+z(t)) = \frac{\dot{S}(t)}{S(t)} + \kappa \nu_{\gamma}(t)
\ . \label{reddiff} \end{equation}
Integration of (\ref{reddiff}) from
$t_0$ to $t_1$ yields
\begin{equation}
1+z_1 = \frac{S(t_1)}{S(t_0)}
\exp ( \int_{t=t_0}^{t=t_1}\kappa \nu_{\gamma}(t)dt )
\label{t0exact} \end{equation} which may be regarded as an implicit equation
for $t_0$ in terms of $z_1$ and $t_1$.  To first order in $\kappa$
equation (\ref{t0exact}) gives
\begin{equation}
 \delta t_0 =
 \frac{1}{H(\tzero)}\int_{t=\tzero}^{t=t_1}\kappa
 \nu_{\smashovercirc{\gamma}}(t)dt + O(\kappa^2) \label{deltat0}
\end{equation}
by means of (\ref{red0}), with $\tzero$ is given implicitly
in terms of $z_1$ and $t_1$ by (\ref{red0}), and with
$\kappa\nu_{\smashovercirc{\gamma}}(t)$
defined by an equation corresponding to (\ref{nudef}) but for
$\overcirc{\gamma}$ in place of $\gamma$:  clearly one has
$\kappa\nu_{\smashovercirc{\gamma}}(t)=\kappa\nu_{\gamma}(t) +
O(\kappa^2)$.  For nearby sources (\ref{deltat0}) gives
\begin{equation} \delta t_0 = \frac{\kappa
\nu_{\smashovercirc{\gamma}}(t_1)}{\overcirc{H}\Dummy^2(t_1)}z_1 +
 O(z_1^2) + O(\kappa^2) \quad \mbox{for $z_1 \ll 1$}
\label{delta0z} \end{equation}
by means of (\ref{red0approx}).

One may define a spatial proper distance of the source from the
observer, with respect to the metric ${\bf g}$, by
\begin{equation}
 s_0 := c\int_{\lambda = \lambda_0 }^{\lambda = \lambda_1 }
(S^2 (t) h_{ij} \frac{dx^i}{d\lambda }
\frac{dx^j}{d\lambda })^{1/2} d\lambda
 = c(t_1 - t_0) \label{s0}
\end{equation}
where (\ref{null1}) has been employed on the right.
For nearby sources (\ref{red0approx}) and (\ref{delta0z})
combine to give
\begin{equation}   t_1-t_0 = \frac{z_1}{\overcirc{H}(t_1)}\bigl( 1 -
\frac{\kappa   \nu_{\smashovercirc{\gamma}}(t_1)}{\overcirc{H}(t_1)}\bigr) +
O(z_1^2) +O(\kappa^2) \quad   \mbox{for $z_1 \ll 1$}  \label{t1t0}
\end{equation}
so, for such sources, the redshift $z_1$ of the source is expressible
as a function of its spatial proper distance $s_0$ according to
\begin{equation}
z_1 (s_0) =
\frac{\overcirc{H}(t_1)s_0}{c}\bigl(
1+\frac{\kappa\nu_{\smashovercirc{\gamma}}(t_1)}{\overcirc{H}(t_1)}\bigr) +
O(s_0^2) + O(\kappa^2) \quad \mbox{for $z_1 \ll 1$} \ .
\label{reddist}
\end{equation}
Notice that in the case $\nu_{\gamma} (t_1)=O(\kappa )$ equation
(\ref{reddist}) agrees,
to first order in $z_1$ and $\kappa$, with the corresponding
formula for the unperturbed FRW space-time.

\section{Luminosity distance and apparent magnitude}
\label{appmag}
Suppose now that, in the perturbed space-time, the source with
world line $\{ {\bf x} = {\bf x}_0\}$ radiates uniformly in all
directions with power $P$.

\newcommand{\boldmbar}{\hspace*{.2ex}\protect\rule[1.5ex]{.6em}{.13ex}\hspace*{-.7em}{\bf m}}
\newcommand{\mbar}{\protect\rule[1.5ex]{.6em}{.13ex}\hspace*{-1.5ex}m}
\newcommand{\mcircbar}{\protect\rule[2.5ex]{.6em}{.13ex}\hspace*{-1.5ex}\overcirc{m}}
Let $k^a:= \frac{dx^a}{d\lambda}$ be the tangent to $\gamma$ and let
${\bf m}$ be a complex vector at $(t_0,{\bf x}_0)$ satisfying
${\bf m}{\cdot}\protect\boldmbar = 1$,
${\bf m}{\cdot} {\bf k} = {\bf m}{\cdot}{\bf m} = {\bf m}{\cdot} {\bf u} =0$.
In order to maintain these conditions along $\gamma$ one requires
that ${\bf m}$ be propagated along $\gamma$ according to \begin{equation}
\nabla_{\bf k}{\bf m} = -\bigl( \frac{{\bf m}{\cdot}
\nabla_{\bf k}{\bf u}}{{\bf u}{\cdot} {\bf k}}\bigr){\bf k} \ .
\label{mprop} \end{equation}  The congruence of all the null geodesics
emanating from $(t_0,{\bf x}_0)$ gives rise to a family of Jacobi fields
along $\gamma$ with expansion $\varrho := k_{a;b}\mbar\dummy^a m
\dummy^b$ and shear $\varsigma := k_{a;b}\mbar\dummy^a\mbar\dummy^b$ with respect to
the 2-frame $\{ {\bf m}, \protect\boldmbar \}$ and the affine parameter
$\lambda$.  Since the congruence is irrotational, the imaginary part
of $\varrho$ is zero.  The standard propagation equations for
$\varrho$ and $\varsigma$ therefore reduce to
\begin{align}
\frac{d}{d\lambda}\varrho & = - \varrho^2 - \varsigma
\overline{\varsigma} - \Theta \label{Ray1}\\
\frac{d}{d\lambda}\varsigma &= -2\varsigma \varrho - \Psi \label{Ray2}
\end{align}
where
\begin{align}
\Theta & :=
\frac{1}{2}R_{ab}k^a k^b =
\frac{1}{2}\kappa T_{ab}k^a k^b \\
\Psi & :=
R_{abcd}k^ak^c\protect\mbar\dummy^b
\protect\mbar\dummy^d = C_{abcd}k^a k^c
\protect\mbar\dummy^b\protect\mbar\dummy^d
\end{align} are the Ricci and Weyl scalars respectively.  (The
fact that ${\bf m}$ is propagated along $\gamma$ according to
(\ref{mprop}), rather than being parallelly propagated as is more usual,
has no effect on the equations (\ref{Ray1}) and (\ref{Ray2}).)

Consider a narrow beam of light rays, centred upon $\gamma$, from the
source point $(t_0,{\bf x}_0)$. Let $\Delta A$ and $I$ be the
cross-sectional area and apparent luminosity of the beam as determined by
comoving observers situated along $\gamma$.  A standard argument from photon
conservation gives that the quantity $(1+z)^2 I\Delta A$ is constant along
$\gamma$. By means of $\varrho = \frac{1}{2}\frac{d}{d\lambda} \ln \Delta A$
one  thus obtains
\begin{equation}
\varrho = \frac{d}{d\lambda} \ln
(I^{-1/2}\frac{dt}{d\lambda}) = \frac{d}{d\lambda} \ln
(I^{-1/2}(1+z)^{-1}) \ .
\end{equation}

In terms of the quantities
\begin{align}
J(\lambda ) & := I^{-1/2}(1+z)^{-1} \\
\xi (\lambda ) & := J^2 (\lambda) \left(
\frac{dt}{d\lambda}\right)^{-1}\varsigma (\lambda )
\label{xidef}
\end{align}
equations (\ref{Ray1}) and (\ref{Ray2}) become
\begin{align}
\frac{d^2}{d\lambda^2}J(\lambda ) +
J^{-3}(\lambda) \xi (\lambda ) \overline{\xi} (\lambda )\left( \frac{dt}{d\lambda}\right)^2
+ J(\lambda )\Theta & =0 \label{Intensity}\\
\frac{d}{d\lambda}\left( \frac{dt}{d\lambda }\xi (\lambda )\right) +  J^2 (\lambda )\Psi & = 0 \ .
\label{Shear1} \end{align}

Substituting for $\lambda$ in terms of $t$ in (\ref{Intensity})
and (\ref{Shear1}) by means of (\ref{d2t}) one thus obtains
\begin{align}
\frac{d^2}{dt^2}J(t) - \left( \frac{\dot{S}(t)}{S(t)}
+\kappa\nu_{\gamma}(t)\right) \frac{d}{dt}J(t) + \xi (t) \bar{\xi}(t)J^{-3}(t)
+ \kappa \tau_{\gamma}(t) J(t) &=0
\label{d2It}\\ \frac{d}{dt}\xi (t) - \left( \frac{\dot{S}(t)}{S(t)}
+\kappa \nu_{\gamma}(t)\right) \xi (t) + J^2(t)\kappa\psi_{\gamma}(t) &=0 \label{Shear2}
\end{align} for \begin{align}
\kappa \tau_{\gamma}(t)  : & \! =
 \frac{1}{2}\kappa T_{ab}\frac{dx^a}{dt}\frac{dx^b}{dt}
 \label{taudef}\\ \kappa \psi_{\gamma} (t) : & \! =
 C_{abcd}\frac{dx^a}{dt}\frac{dx^c}{dt} \protect\mbar\dummy^b
\protect\mbar\dummy^d  \label{littlepsidef} \\ & =
2E_{ij}\mbar\dummy^i\mbar\dummy^j +
\frac{2}{S^2(t)}\prethree\eta^k\dummy_{ij}B_{lk}\mbar\dummy^j\mbar\dummy^l\frac{dx^i}{dt}
\label{psiEB} \ . \end{align}
The initial conditions for (\ref{d2It}) and (\ref{Shear2}) are
\begin{equation} \left. \begin{aligned}
\xi (\tzero ) & =0\\
J(\tzero ) & =0\\
\left. \frac{d}{dt}\right|_{t=\smashovercirc{t}_0}J(t) & =
c\sqrt{\frac{4\pi}{P}}
\end{aligned} \quad \right\}   \ . \label{IC}
\end{equation}

To first order in $\kappa$, the solutions to (\ref{d2It}) and (\ref{Shear2}),
subject to the initial conditions
(\ref{IC}), are
\begin{align}
J(t) & = \sqrt{\frac{4\pi}{P}}\frac{c}{S(t_0)}\int_{t'=t_0}^{t'=t}S(t')(1+\kappa\mu_{\smashovercirc{\gamma}}(t'))dt'
+ O(\kappa^2) \label{Trial} \\
\xi (t) & = -\frac{4\pi c^2}{P}\frac{S(t)}{S(t_0)}\int_{t'=t_0}^{t'=t}
\frac{\Sigma^2(t_0,t')}{S(t')}\kappa\psi_{\smashovercirc{\gamma}} (t')dt' +
O(\kappa^2)\label{xisoln} \end{align} for
\begin{equation}
\Sigma (t_0, t) := \int_{t'=t_0}^{t'=t}S(t')dt'  \label{sigmadef}
\end{equation}
and \begin{equation}
\kappa \mu_{\smashovercirc{\gamma}}(t) :=
\int_{t'=t_0}^{t'=t}(\kappa\nu_{\smashovercirc{\gamma}}(t') - \frac{\Sigma
(t_0,t')}{S(t')}\kappa\tau_{\smashovercirc{\gamma}}(t'))dt' \label{mut}
\end{equation}
where $\kappa\tau_{\smashovercirc{\gamma}}(t)$ and $\kappa \psi_{\smashovercirc{\gamma}}(t)$ are defined by equations
analogous to (\ref{taudef}) and (\ref{littlepsidef}) for
$\overcirc{\gamma}$ in place of $\gamma$. Clearly one has $\kappa
\tau_{\smashovercirc{\gamma}}(t) = \kappa\tau_{\gamma}(t)
+O(\kappa^2)$ and $\kappa \psi_{\smashovercirc{\gamma}}(t) =
\kappa\psi_{\gamma}(t) + O(\kappa^2)$.

From (\ref{xidef}), (\ref{Trial}) and (\ref{xisoln}) one obtains
\begin{equation} \varsigma (t)  = -
\frac{S(\overcirc{t}_0)S(t)}{\Sigma^2(\overcirc{t}_0,t
)}\frac{dt}{d\lambda}\int_{t'=\overcirc{t}_0}^{t'=t}\frac{\Sigma^2(\overcirc{t}_0,t')}{S(t')}\kappa
\psi_{\smashovercirc{\gamma}}(t')dt' + O(\kappa^2)
\label{shear3}
\end{equation} for the shear as a function of $t$.

Writing $t_0=\tzero +\delta t_0$  and setting $t=t_1$ in (\ref{Trial}) one obtains
\begin{multline}
I^{-1/2}(z_1) =  c\sqrt{\frac{4\pi}{P}}\frac{(1+z_1)^2}{S(t_1)}\{
\Sigma(\tzero,t_1)
\bigl( \frac{S^2(\tzero)}{\dot{S}(\tzero)}
\Sigma(\tzero,t_1)\bigr)
\int_{t=\smashovercirc{t}_0}^{t=t_1}\kappa\nu_{\smashovercirc{\gamma}}(t)dt \\
+\int_{t=\smashovercirc{t}_0}^{t=t_1}S(t)\kappa\mu_{\smashovercirc{\gamma}}(t)dt
\} +O(\kappa^2)\end{multline} by means of (\ref{deltat0}) and
\ref{sigmadef}), where $\overcirc{t}\dummy_0$ is given in terms
of $t_1$ and $z_1$ by (\ref{red0}).  From this one obtains
\begin{multline}
I^{-1/2}(z_1) =c\sqrt{\frac{4\pi}{P}}\frac{(1+z_1)^2}{S(t_1)}
\{ \Sigma (\overcirc{t}\dummy_0,t_1) -
\frac{S^2(\overcirc{t}\dummy_0)}{\dot{S}(t_1)}
\int_{t=\smashovercirc{t}\dummy_0}^{t=t_1}
\kappa\nu_{\smashovercirc{\gamma}}(t)dt
 \\[2ex]
 -\int_{t=\smashovercirc{t}\dummy_0}^{t=t_1}\Sigma
(\overcirc{t}\dummy_0,t)(\kappa\nu_{\smashovercirc{\gamma}}(t)  -
\frac{\Sigma
(\overcirc{t}\dummy_0,t)}{S(t)}\kappa\tau_{\smashovercirc{\gamma}}(t))dt
\\ -\Sigma(\overcirc{t}\dummy_0,t_1)
\int_{t=\smashovercirc{t}\dummy_0}^{t=t_1} \frac{\Sigma
(\overcirc{t}\dummy_0,t)}{S(t)}\kappa\tau_{\smashovercirc{\gamma}}(t)dt\}
+O(\kappa^2) \label{It1z}
\end{multline}
by means of (\ref{mut}), (\ref{sigmadef}) and an
integration by parts. For nearby sources (\ref{It1z}) gives
\begin{equation}
I^{-1/2}(z_1) =
\sqrt{\frac{4\pi}{P}}\frac{cz_1}{\overcirc{H}(t_1)}\bigl(
1-\frac{\kappa\nu_{\smashovercirc{\gamma}}(t_1)}{\overcirc{H}(t_1)}\bigr)
+ O(z_1^2) + O(\kappa^2) \quad \mbox{for $z_1 \ll 1$}
\label{It1zapprox}
\end{equation} by means of (\ref{t1t0}).  The equation corresponding to (\ref{It1z}) for the unperturbed
metric $\overcirc{\bf g}$ is
\begin{equation}
\overcirc{I}\Dummy^{-1/2}(z_1) =
c\sqrt{\frac{4\pi}{P}}\frac{(1+z_1)^2}{S(t_1)}\Sigma(\overcirc{t}\dummy_0,t_1)
\ . \label{It1zbg}
\end{equation}

From (\ref{shear3}), the total shear of the image at the
observation point $(t_1,{\bf x}_1)$ is given by
\begin{equation}  \int_{\lambda = \lambda_0}^{\lambda = \lambda_1}
\varsigma d\lambda =
-\int_{t=\overcirc{t}_0}^{t=t_1}\frac{S(\overcirc{t}_0)S(t)}{\Sigma^2(\overcirc{t}_0,t)}\left(
\int_{t'=\overcirc{t}_0}^{t'=t}\frac{\Sigma^2(\overcirc{t}_0,t')}{S(t')}\kappa\psi_{\smashovercirc{\gamma}}(t')dt'\right)
dt + O(\kappa^2) \ . \label{shear4} \end{equation}

The luminosity distance of the source is defined by
\begin{equation}
d_L := \sqrt{\frac{P}{4\pi I(z_1)}}  \label{lumdist}
\end{equation}
which, for nearby sources, combines with (\ref{It1zapprox}) to
give
\begin{equation}
d_L = \frac{cz_1}{\overcirc{H}(t_1)}\bigl(
1-\frac{\kappa\nu_{\smashovercirc{\gamma}}(t_1)}{\overcirc{H}(t_1)}\bigr)
+ O(z_1^2) + O(\kappa^2) \quad \mbox{for $z_1 \ll 1$} \ .
\label{dLzapprox}
\end{equation}
Note that in the case $\nu_{\gamma} (t_1) =
O(\kappa )$ this agrees, to first order in $z_1$ and $\kappa$,
with the corresponding formula for the unperturbed FRW space-time.
From (\ref{dLzapprox}) and (\ref{reddist}) one has
\begin{equation} d_L = s_0 + O(s_0^2) + O(\kappa^2)
\label{lumdistapprox}
\end{equation}
which gives the acceptable result that the spatial proper distance
and luminosity
distance agree to first order in $s_0$ and $\kappa$,
independent of the value of $\nu_{\smashovercirc{\gamma}}(t_1)$.  This
holds
for both the perturbed metric ${\bf g}$ and the unperturbed metric
$\overcirc{\bf g}$.

\newcommand{\rf}{\mbox{\protect\scriptsize \em ref}}
The apparent magnitude of the source at ${\bf x} = {\bf x_0}$
relative to a reference source at ${\bf x} = {\bf
x}_{\protect\rf}$ is defined by
\begin{equation}
m := m_{\protect\rf} + \frac{5}{2}\log_{10}
\frac{I_{\protect\rf}}{I} \label{mdef}
\end{equation}
where $I$ and $I_{\protect\rf}$ are the apparent luminosities of
the respective sources.  The reference source is taken to have
power $P_{\protect\rf}$ and to be at a spatial proper distance
$s_{0,\protect\rf}$ that is small on a cosmological scale (10
parsecs is conventional).  For the unperturbed metric
$\overcirc{\bf g}$, a source of apparent luminosity $\overcirc{I}$
has an apparent magnitude
\begin{equation} \overcirc{m} = \overcirc{m}_{\protect\rf} +
\frac{5}{2} \log_{10}
\frac{\overcirc{I}_{\protect\rf}}{\overcirc{I}} \label{mdefbg}
\end{equation}
relative to a source of apparent luminosity
$\overcirc{I}_{\protect\rf}$.  The sources will again be taken to
have powers $P$ and $P_{\protect\rf}$ respectively, with the
reference source at the same spatial proper distance $s_{0,\protect\rf}$
as for the perturbed metric ${\bf g}$.  Let
$m_{\protect\rf}=\overcirc{m}_{\protect\rf}$.  The objective
sources for ${\bf g}$ and $\overcirc{\bf g}$ are both taken to be
at redshift $z_1$.  From (\ref{mdef}) and (\ref{mdefbg}) one
obtains
\begin{equation}
\delta m(z_1) = \frac{5}{2\ln 10} \bigl\{ \lim_{s\rightarrow 0}
\ln \frac{I_{\protect\rf}(s)}{\overcirc{I}_{\protect\rf}(s)} - \ln
\frac{I(z_1)}{\overset{\circ}{I}(z_1)} \bigr\}
\end{equation}
in the limit $s_{0,\protect\rf} \rightarrow 0$.  By
(\ref{lumdistapprox}), the first term in the braces is just
$\ln(P/P_{\protect\rf})$ to first order in $\kappa$. By
(\ref{It1z}) and (\ref{It1zbg}) one thus obtains
\begin{multline}
\delta m(z_1) = \frac{5}{\ln 10}\{ -
\frac{S^2(\tzero)}{\dot{S}(t_1)\Sigma (\overset{\circ}{t}_0 ,t_1)}
\int_{t=\smashovercirc{t}_0}^{t=t_1}\kappa
\nu_{\smashovercirc{\gamma}}(t)dt
-\int_{t=\smashovercirc{t}_0}^{t=t_1} \frac{\Sigma
(\tzero,t)}{S(t)}\kappa\delta\tau_{\smashovercirc{\gamma}}(t)dt
 \\ - \frac{1}{\Sigma(
\tzero,t_1)}\int_{t=\smashovercirc{t}_0}^{t=t_1}\Sigma
(\overset{\circ}{t}_0,t)(\kappa\nu_{\smashovercirc{\gamma}}(t) -
\frac{\Sigma
(\tzero,t)}{S(t)}\kappa\delta\tau_{\smashovercirc{\gamma}}(t))dt
\} + O(\kappa^2) \label{deltamz}
\end{multline}
for the correction to the apparent magnitude-redshift relation.
Note that this correction depends on the perturbation $\delta {\bf
h}$ only through the functions
$\kappa\nu_{\smashovercirc{\gamma}}(t)$ and
$\kappa\delta\tau_{\smashovercirc{\gamma}}(t)$.

For nearby sources (\ref{deltamz}) gives
\begin{equation}
\delta m(z_1) = -\frac{5}{\ln
10}\frac{\kappa\nu_{\smashovercirc{\gamma}}(t_1)}{\overset{\circ}{H}(t_1)}
+ O(z_1) + O(\kappa^2) \quad \mbox{for $z\ll 1$}
\label{deltamzapprox}
\end{equation}
for the apparent magnitude-redshift correction.  This evidently
vanishes at $z_1=0$ to first order in $\kappa$ in the case $
\nu_{\gamma}(t_1)=O(\kappa )$.

It is evident from the analysis in the present and preceding
sections that the equation
\begin{equation} \kappa\nu_{\gamma}(t_1) = 0 +O(\kappa^2 ) \label{nufix}
\end{equation}
may be interpreted as a condition on the perturbation which, as
regards local optical properties, ensures that the background FRW
metric $\overcirc{\bf g}$ is a best fit to the perturbed metric
${\bf g}$, in the direction of $\dot{\gamma}(t_1)$, at the
observation point $(t_1,{\bf x}_1)$.  A requirement that (\ref{nufix})
holds for all null geodesics $\gamma$ through $(t_1,{\bf x}_1)$ is, by
(\ref{nudef}), equivalent to the condition
\begin{equation} \partial_t \delta h_{ij} (t_1,{\bf x}_1) =0 +O(\kappa^2) \ .
\label{newgauge} \end{equation}
In general this represent a physical
constraint on the perturbation at $(t_1,{\bf x}_1)$ since there is, in
general, no freedom to choose a new time slicing such that the $3$-metric
${\bf h}$ is intrinsic to the level surfaces of the new time.  However, for
the specific class of perturbations to be introduced in \S\ref{Newt}, there is
just such a  freedom, at least to first order in $\kappa$, which may be
exploited to  ensure that (\ref{newgauge}) does hold.

\section{Newtonian perturbations} \label{Newt}
In order to describe the matter distribution of the cosmos,
physical considerations suggest that one seeks a perturbed
$3$-metric ${\bf h}$ such that the corresponding space-time metric
${\bf g}$ has an energy tensor of the form
\begin{equation}
T^{ab} = (\rho +p )u^au^b + pg^{ab} \label{Tpf}
\end{equation}
describing a perfect fluid with velocity $u^a:= t^{;a}$,
density
 $\rho$ and pressure $p$. It is conventional to define a
dimensionless density parameter by
\begin{equation} \Omega:= \frac{\kappa \rho}{3H^2} \ .
\label{Omegadef}
\end{equation} Note that all of $\rho$, $H$ and $\Omega$ depend on
all four coordinates.  With regard to the pressure, since the form
of (\ref{gdef}) implies that $u^a$ is geodesic, the conservation
equation for (\ref{Tpf}) implies $h_a\Dummy^bp_{;b}=0$ and hence
$h_a\Dummy^b(\delta p)_{;b}=0$ under the assumption that $\rho +
p$ is nowhere zero. This leads one to consider the particular case
$\delta p =0$ which describes perturbations arising from the
addition or removal of comoving dust.

The full non-linear problem presents formidable difficulties,
although one does have by (\ref{Tpf}) and (\ref{Ritg})  that
$\delta h_{ij}$ satisfies the simple equation
\begin{equation}
h^{jk}(\partial_t\delta h\Dummy_{k[i})\Dummy_{\protect\bbar j]} =0
\ . \label{hsimp}\end{equation} In order to make progress, only a
linear approximation to a solution will be sought.  Specifically
the problem is to obtain $\delta h_{ij}$, regarded as a power
series in $\kappa$, vanishing to zeroth order, such that the
energy tensor of $g_{ab}$ has the form
\begin{equation}
T^{ab} = (\rho + p)u^au^b + pg^{ab} + O(\kappa ) \ .
\label{TabMcV}\end{equation}

Consider a perturbation of the $3$-metric $\overcirc{h}_{ij}$ of
the form
\begin{equation}
\delta h_{ij} = F(t)\kappa \Phi_{\bbar ij} + G(t)\kappa \Phi
h_{ij} +O(\kappa^2 ) \label{McVdeltah}
\end{equation} for functions $F(t)$, $G(t)$ and a
$t$-independent scalar field $\Phi ({\bf x})$. The form of
(\ref{McVdeltah}) corresponds to the synchronous gauge perturbations
considered by Mukhanov et al.\ (1992, p.216).  It will be
convenient, although not necessary, to regard $F(t)$ and $G(t)$ as
power series in $\kappa$ even though only the zeroth order terms
will be of significance. Note that although $\delta h_{ij}$ enters
into the right side of (\ref{McVdeltah}) through the term
$\Phi_{\protect\bbar ij}$, this is not significant in the linear
approximation since $\Phi_{\protect\bbar ij}$ and
$\Phi_{\smashovercirc{\bbar}ij}$ agree to zeroth order in
$\kappa$.

\newcommand{\bracketdot}{\! \raisebox{1.7ex}{.}\: }
From first principles one has
\begin{equation}
\delta \prethree R_{ij}-2k\delta h_{ik} = -\frac{1}{2}
G(t)\{\kappa \Phi _{\bbar ij} + ( \kappa \prethree\Delta \Phi  +
4k\kappa \Phi )h_{ij}\} + O(\kappa^2) \label{delta3Rijphi}
\end{equation}
which by (\ref{delta3R1}) gives
\begin{equation}\delta\prethree R = -2G(t)(\kappa\prethree\Delta  \Phi +
3k\kappa \Phi ) + O(\kappa^2) \ .  \label{delta3Rphi}
\end{equation}

Substitution of (\ref{McVdeltah}) into (\ref{Tgtt}), (\ref{Tgit})
and (\ref{Tgij}) yields
\begin{align} \kappa T_{tt} & =\kappa \overcirc{\rho} + \bigl(
\frac{\dot{S}(t)\dot{F}(t)}{S^2(t)} - \frac{G(t)}{S^2(t)} \bigr)
\kappa\prethree\Delta  \Phi + 3 \bigl(
\frac{\dot{S}(t)\dot{G}(t)}{S(t)} - \frac{kG(t)}{S^2(t)}\bigr)
\kappa \Phi +O(\kappa^2) \label{Tttphi}\\ \kappa T_{it} &= \kappa
T_{ti} = (k\dot{F}(t) - \dot{G}(t))\kappa \Phi_{\protect\bbar i} +
O(\kappa^2) \label{Titphi} \\ \kappa T_{ij} &= \{ S^2(t) \kappa
\overcirc{p} + \frac{1}{2} (G(t) -
\frac{1}{S(t)}(S^3(t)\dot{F}(t))\bracketdot )\kappa
\prethree\Delta \Phi \nonumber \\
 & \quad +(kG(t) - \frac{1}{S(t)}(S^3(t)\dot{G}(t)) \bracketdot )\kappa \Phi
 \} h_{ij} \nonumber \\
 & \quad + \frac{1}{2} \bigl( \frac{1}{S(t)}(S^3(t)\dot{F}(t))\bracketdot
 - G(t)\bigr) \kappa \Phi_{\protect\bbar ij} + O(\kappa^2)  \label{Tijphi}
\end{align} by means of (\ref{delta3Rphi}), (\ref{3Riccih0}) and
(\ref{delta3Rijphi}).

For $\kappa T_{it}$ and the trace-free part of $\kappa T_{ij}$ to
vanish to first order in $\kappa$, in accordance with
(\ref{TabMcV}), it suffices to require that $F(t)$ and $G(t)$
satisfy
\begin{align} G(t) & =
\frac{1}{S(t)}(S^3(t)\dot{F}(t))\bracketdot + O(\kappa )
\label{Gdef} \\ \dot{G}(t) & = k\dot{F}(t) + O(\kappa ) \ .
\label{Gdot}
\end{align}
From these one obtains
\begin{equation} (S(t)(S^2(t)\dot{F}(t))\bracketdot )\bracketdot = 0 + O(\kappa
) \end{equation} by means of (\ref{Hdot0}), and hence
\begin{equation} S(t)(S^2(t)\dot{F}(t))\bracketdot = C + O(\kappa
) \label{FC} \end{equation} for some constant $C$.

If $C$ were chosen to vanish to zeroth order in $\kappa$ then, by
(\ref{FC}) and (\ref{Gdef}), both $(S^2(t)\dot{F}(t))\bracketdot$
and $G(t) - \dot{S}(t)S(t)\dot{F}(t)$ would vanish to zeroth order
in $\kappa$ and so, as will be evident from (\ref{gdiag}) in \S
\ref{gauge}, ${\bf g}$ would be an FRW metric to first order in
$\kappa$ irrespective of the function $\Phi ({\bf x})$. To avoid
this uninteresting case $C$ will be chosen to be non-zero to
zeroth order in $\kappa$. One may then normalise $F(t)$ to give
\begin{equation} C = 1 \ . \label{C1} \end{equation}
Equations (\ref{Gdef}), (\ref{FC}) and (\ref{C1}) give
\begin{align} (S^2(t)\dot{F}(t))\bracketdot & = \frac{1}{S(t)} +
O(\kappa ) \label{Fdd} \\ G(t) - \dot{S}(t)S(t)\dot{F}(t) & =
\frac{1}{S(t)} + O(\kappa ) \ . \label{G} \end{align} These are
equivalent to equations (\ref{Gdef}), (\ref{Gdot}) and (\ref{C1})
by virtue of (\ref{Hdot0}).

By means of (\ref{Fdd}) and (\ref{G}), equations (\ref{Tttphi}),
(\ref{Titphi}) and (\ref{Tijphi}) combine to give that $T^{ab}$
has the required form (\ref{TabMcV}) for
\begin{align} \delta \rho & = - \frac{1}{S^3(t)}(
\prethree\Delta \Phi + 3k \Phi ) + O(\kappa) \label{McVdeltarho}\\
\delta p &= 0 + O(\kappa) \ . \label{McVdeltap} \end{align} From
(\ref{McVdeltarho}) one has \begin{equation}
\partial_t(S^3(t)\delta \rho ) = 0 + O(\kappa) \label{deltarhoMcVt}\end{equation}
as one would expect since the perturbation has a dust equation of
state.

From (\ref{McVdeltah}), (\ref{Gdot}) and (\ref{McVdeltarho}) one
has
\begin{equation} h^{ij}\partial_t\delta h_{ij} =
-S^3(t)\dot{F}(t)\kappa \delta \rho +O(\kappa^2) \end{equation}
whereby (\ref{deltaexp}), (\ref{deltaq}) and (\ref{delta3R}) give
\begin{align}
\delta H & = -\frac{1}{6}S^3(t)\dot{F}(t)\kappa \delta \rho
+O(\kappa^2) \label{deltaexpMcV} \\
\delta q & = \frac{1}{2}\{ S(t)(S^2(t)\dot{F}(t))\bracketdot
+ 2S^3(t)\dot{S}(t)\dot{F}(t)(1+\overcirc{q}(t))\}\delta \Omega +O(\kappa^2)
\label{deltaqagain} \\
\delta \prethree R & =
2S^3(t)G(t)\kappa \delta \rho +O(\kappa^2)\label{delta3RMcV}
\end{align} with the help of (\ref{Omegadef}) and (\ref{G}).

It is straightforward to check, by means of (\ref{Hijg}) and
(\ref{3Riccih0}), that the perturbed space-time metric {\bf g}
defined by (\ref{gdef}), (\ref{hdef}) and (\ref{McVdeltah}) is
silent to first order in $\kappa$ in the sense of
\begin{equation} B_{ij} = 0 + O(\kappa^2
) \ . \label{silentMcV}
\end{equation}
The electric part of the Weyl tensor of ${\bf g}$ is, by
(\ref{Eijg}), (\ref{delta3Rijphi}), (\ref{delta3Rphi}),
(\ref{Fdd}) and (\ref{G}), given by
\begin{equation}
E_{ij} = -\frac{1}{2S(t)} (\kappa \Phi_{\protect\bbar ij} -
\frac{1}{3} h_{ij}\kappa\prethree\Delta \Phi ) + O(\kappa^2) \ .
\label{EMcV} \end{equation}

By (\ref{TabMcV}) and the analogue of (\ref{taudef}) for
$\overcirc{\gamma}$, the function $\delta
\tau_{\smashovercirc{\gamma}}(t)$ in (\ref{deltamz}) is given by
\begin{equation} \delta \tau_{\smashovercirc{\gamma}}(t) =
\frac{1}{2}\delta \rho (\overcirc{\gamma}(t)) + O(\kappa ) \ .
\label{deltatau}
\end{equation}  An equation for the function $\nu_{\smashovercirc{\gamma}}(t)$
in (\ref{deltamz}) will be given in {\S}\ref{gauge}. By
(\ref{silentMcV}), (\ref{EMcV}) and the analogue of (\ref{psiEB})
for $\overcirc{\gamma}$ the function
$\psi_{\smashovercirc{\gamma}} (t)$ in equation (\ref{shear4}) is
given by
\begin{equation} \psi_{\smashovercirc{\gamma}}(t) =
-\frac{1}{S(t)} \Phi_{\smashovercirc{\protect\bbar}
ij}\mcircbar\dummy^i \mcircbar\dummy^j + O(\kappa )
\end{equation} where $\overcirc{m}\dummy^i$ is the analogue
of $m^i$ for $\overcirc{\gamma}$.

Space-times satisfying the condition $B_{ab}=0$ have been termed
`silent' by some authors and `Newtonian-like' by others. However
it is known that the vanishing of $B_{ab}$ can be conserved in
time only in specialised cases (Maartens {\it et al.} 1998).  In
any second or higher order study of perturbations of FRW
cosmologies one would therefore not expect $B_{ab}$ to vanish to
any higher than first order in $\kappa$.

\section{The gauge condition} \label{gauge}
The functions $F(t)$ and $G(t)$, which enter explicitly into the
perturbed metric ${\bf g}$, are not determined uniquely by
(\ref{Fdd}) and (\ref{G}) since these equations allow two freely
specifiable constants of integration. In alternative terminology,
there is a gauge freedom
\begin{align} F(t) & \rightarrow F(t) +  A + A_0\int_{t'=t}^{t'=t_1}
\frac{dt'}{S^2(t)}  \label{Ftrans} \\ G(t) & \rightarrow G(t) -
A_0\frac{\dot{S}(t)}{S(t)} \label{Gtrans} \end{align} where $A_0$
and $A$ are real constants.  In order to understand the meaning of
this freedom, consider the metric ${\bf g}$, as given by
(\ref{gdef}), (\ref{hdef}) and (\ref{McVdeltah}), expressed with
respect to the coordinate system employed in (\ref{hFRW}):
\begin{multline}
{\bf g} = -dt^2 + S^2(t) \left(
1+\frac{kr^2}{4}\right)^{-2}(1+G(t)\kappa \Phi ({\bf x})
)\sum_{i=1}^3 (dx^i)^2
\\ +S^2(t)F(t)\kappa \Phi ({\bf x})_{\protect\bbar ij}dx^idx^j + O(\kappa^2)
\ . \label{gccord}
\end{multline} In terms of new coordinates $(\tilde{t}\/,\tilde{\bf
x})$ defined by
\begin{align} \tilde{t} & := t + \frac{1}{2} F\highstar (t)\kappa
\Phi ({\bf x}) \\ \tilde{\bf x}\dummy^i & := x^i + \frac{1}{2}
F(t)\left( 1+\frac{kr^2}{4}\right)^2\kappa \partial_i\Phi({\bf x})
\ ,
\end{align}
for an as yet unspecified function $F\highstar (t)$, the metric
${\bf g}$ assumes the form
\begin{multline} {\bf g} = -(1-\dot{F}\highstar (\tilde{t}\/)\kappa
\Phi (\tilde{\bf x}))d\tilde{t}\/ \dummy^2 + (F\highstar
(\tilde{t}\/)-S^2(\tilde{t}\/)\dot{F}(\tilde{t}\/))\kappa
\partial_{\tilde{\bf x}\dummy^i}\! \Phi (\tilde{\bf x})\,
d\tilde{x}\dummy^id\tilde{t}\\ + \frac{S^2(\tilde{t}\/)}{ \left(
1+\frac{k \tilde{r}\dummy^2}{4}\right)^2}\bigl( 1+(G(\tilde{t}\/)
-\frac{\dot{S}(\tilde{t}\/)}{S(\tilde{t}\/)} F\highstar
(\tilde{t}\/))\kappa \Phi (\tilde{\bf x})\bigr) \sum_{i=1}^3
(d\tilde{x}\dummy^i )^2 + O(\kappa^2 ) \ .
\end{multline}
With the choice \begin{equation} F\highstar (t) = S^2
(t)\dot{F}(t)
\end{equation}  the metric ${\bf g}$ assumes the diagonal form
\begin{multline} {\bf g} = -
(1-(S^2(\tilde{t})\dot{F}(\tilde{t}\/))\bracketdot\kappa \Phi
(\tilde{\bf x}))d\tilde{t}\/\dummy^2 \\ +
\frac{S^2(\tilde{t}\/)}{\left(
1+\frac{k\tilde{r}\dummy^2}{4}\right)^2}
(1+(G(\tilde{t}\/)-\dot{S}(\tilde{t}\/)S(\tilde{t}\/)\dot{F}(\tilde{t}\/))\kappa
\Phi (\tilde{\bf x}))\sum_{i=1}^3 (d\tilde{x}\dummy^i)^2
+O(\kappa^2) \ . \label{gdiag}\end{multline}

No assumptions about the $t$-dependence of $F(t)$ and $G(t)$ have
yet been employed. It is evident that if
$(S^2(t)\dot{F}(t))\bracketdot$ and
$G(t)-\dot{S}(t)S(t)\dot{F}(t)$ were both to vanish to zeroth
order in $\kappa$ then, as quoted in \S\ref{Newt}, ${\bf g}$ would
be an FRW metric to first order in $\kappa$.

For $F(t)$ and $G(t)$ satisfying (\ref{Fdd}) and (\ref{G}),
equation (\ref{gdiag}) reduces to
\begin{equation} {\bf g} = -\left( 1-\frac{\kappa \Phi(\tilde{\bf
x})}{S(\tilde{t})}\right) d\tilde{t}\dummy^2 +
\frac{S^2(\tilde{t})}{(1+\frac{k\tilde{r}\dummy^2}{4})^2}\left(
1+\frac{\kappa \Phi(\tilde{\bf x})}{S(\tilde{t})}\right)
\sum_{i=1}^3(d\tilde{ x}\dummy^i)^2 +O(\kappa^2) \label{gMcV}
\end{equation} which is the metric of Newman \& McVittie (1982).
Note that the curves $\{ \tilde{\bf x} = \mbox{\em const.}\}$ have unit
tangent
\begin{equation} \tilde{u}_a = -\left( 1+\frac{\kappa \Phi(\tilde{\bf
x})}{2S(\tilde{t})}\right) \tilde{t}_{\tilde{;}a} + O(\kappa^2)
\end{equation} and so are non-geodesic to first order in $\kappa$.
They are therefore not the world lines of freely falling
observers.

The absence of $F(t)$ and $G(t)$ in (\ref{gMcV}) shows that the isometry class
of ${\bf g}$
 is unaffected by the choice of integration constants for
(\ref{Fdd}) and (\ref{G}).  Indeed the gauge transformation
(\ref{Ftrans}), (\ref{Gtrans}) is induced by the coordinate
transformation
\begin{align} t & \rightarrow t + \frac{A_{0}}{2}\kappa \Phi ({\bf
x}) \label{ttrans} \\ x^i & \rightarrow x^i - \frac{1}{2}\bigl(
A+A_0\int_{t'=t}^{t'=t_1}\frac{dt'}{S^2(t')}\bigr)
\left(1+\frac{kr^2}{4}\right)^2\kappa \partial_i \Phi ({\bf x})
 \label{xtrans}
\end{align}
which preserves the form of the metric (\ref{gccord}) for $F(t)$, $G(t)$
transforming according to (\ref{Ftrans}) and (\ref{Gtrans}).
Moreover the transformation (\ref{ttrans}) of $t$ preserves the
form of (\ref{McVdeltah}) on the level surfaces of $t$, for a given
$\Phi$ on the space-time manifold, for $F(t)$, $G(t)$ transforming according
to (\ref{Ftrans}) and (\ref{Gtrans}).  The gauge freedom (\ref{Ftrans}),
(\ref{Gtrans}) is thus to be interpreted as a freedom to choose the time
function $t$ without violating the form of (\ref{McVdeltah}).  The coordinate
freedom (\ref{ttrans}), (\ref{xtrans}) corresponds to that identified by
Mukhanov et al.\ (1992, p.216).

Although the integration constants for (\ref{Fdd}) and (\ref{G})
do not affect the isometry class of ${\bf g}$ to
first order in $\kappa$, they nonetheless carry physical
significance since they help determine the physically significant
coordinate $t$. In order to make physically appropriate choices of
these constants, consider the function
$\nu_{\gamma}(t)$ of (\ref{nudef}) along a light
ray $\gamma$ through the observation point $(t_1,{\bf x}_1)$.  By
means of (\ref{McVdeltah}), (\ref{Gdot}) and the null geodesic
equations (\ref{null1}) and (\ref{geodesic}) one obtains
\begin{equation}  \nu_{\gamma}(t) =
\dot{F}(t)\{ S(t)\frac{d}{dt}\left( S(t)\frac{d}{dt} \Phi (\gamma
(t))\right) +  \Phi (\gamma (t)) \} + O(\kappa)\ . \label{nuMcV}
\end{equation}
By (\ref{McVdeltah}) and (\ref{Gdot}) the gauge condition (\ref{newgauge})
becomes
\begin{equation} \dot{F}(t_1) = 0 \
. \label{FMcVIC}  \end{equation}
This holds for a unique value of $A_0$ in (\ref{Ftrans}), (\ref{Gtrans}).

Equations (\ref{Fdd}), (\ref{G}) and (\ref{FMcVIC}) give
\begin{align} \dot{F}(t) & =
-\frac{1}{S^2(t)}\int_{t'=t}^{t'=t_1}\frac{dt'}{S(t')} +O(\kappa)
\label{Fdotfix} \\ G(t) & = \frac{1}{S(t)} -
\frac{\dot{S}(t)}{S(t)}\int_{t'=t}^{t'=t_1}\frac{dt'}{S(t')} +
O(\kappa) \ . \label{Gfix} \end{align} Thus, to zeroth order in
$\kappa$, $G(t)$ is now uniquely specified whilst $F(t)$ is
determined only up to an arbitrary additive constant. By
(\ref{ttrans}) and (\ref{xtrans}) this remaining gauge freedom
corresponds to a coordinate transformation of the form
\begin{align} t & \rightarrow t \\
x^i & \rightarrow x^i - \frac{A}{2}\left(
1+\frac{kr^2}{4}\right)^2\kappa
\partial_i\Phi ({\bf x}) \end{align} where $A$ is an arbitrary
constant.  Since this preserves the level surfaces of $t$ it has
no significance for the physical properties of the perturbed
cosmology.  There is therefore no need to specify the remaining
integration constant of (\ref{Fdd}) and (\ref{G}).

By (\ref{deltamzapprox}) the gauge fixing condition (\ref{FMcVIC}) may
be interpreted as a necessary and sufficient condition that there is a
vanishing correction to the apparent magnitude-redshift relation,
in all directions, to zeroth order in the redshift and first order
in $\kappa$.  By (\ref{dLzapprox}) another interpretation is that there
is a vanishing correction to the luminosity distance-redshift relation,
in all directions, to zeroth order in the redshift and first order in
$\kappa$.

An immediate consequence of the gauge fixing condition
(\ref{FMcVIC}) is, by (\ref{deltaexpMcV}), that the Hubble expansion
$H$ of the perturbed
cosmology satisfies
\begin{equation} H(t_1,{\bf x}) = \overcirc{H}(t_1)+O(\kappa^2)
\end{equation} and so is unperturbed to first order in $\kappa$
at all points of the surface $\{ t =t_1\}$. By the use of (\ref{FC}) and
(\ref{FMcVIC}) in (\ref{deltaqagain}) one also has
\begin{equation} \delta q = \frac{1}{2} \delta \Omega + O(\kappa^2) \ .
\label{deltayetagain} \end{equation}
The gauge condition (\ref{FMcVIC}) thus ensures that the perturbation
(\ref{McVdeltah}) affects the deceleration parameter but not the Hubble
parameter at the observation point $(t_1,{\bf x}_1)$.

Subject to the gauge fixing condition (\ref{FMcVIC}) the apparent
magnitude-redshift relation is given by (\ref{deltamz}), with
$\nu_{\smashovercirc{\gamma}}(t)$ given by (\ref{nuMcV}) and
(\ref{Fdotfix}) and $\delta\tau_{\smashovercirc{\gamma}}(t)$ by
(\ref{deltatau}).  Substitution for these quantities in
(\ref{deltamz}) yields
\begin{align} \delta m(z_1) &= \nonumber \\ &
\hspace*{-3em}\frac{5}{\ln 10} \Bigl\{ \frac{S^4(\tzero )
\dot{F}(\tzero )}{2\dot{S}(t_1)\Sigma(\tzero ,t_1)}\left.
\frac{d}{dt} \right|_{t=\tzero }\kappa\Phi(\overcirc{\gamma} (t))
+ \frac{S^2(\tzero )}{\dot{S}(t_1)\Sigma (\tzero ,t_1
)}\int_{t=\tzero }^{t=t_1} \frac{\dot{S}(t)}{S^2(t)}\kappa \Phi
(\overcirc{\gamma} (t))dt\Bigr. \nonumber
\\ &  +\frac{S^2(\tzero )}{\dot{S}(t_1)\Sigma(\tzero , t_1)} \left[
\bigl( \frac{1}{S(t)} - \frac{1}{2}G(t)\bigr) \kappa \Phi
(\overcirc{\gamma}(t))\right]^{t=t_1}_{t=\tzero}  +\frac{\kappa
\Phi (\overcirc{\gamma} (t_1))}{2S(t_1)} \nonumber \\ &  -
\frac{1}{\Sigma (\tzero ,t_1)}\int_{t=\tzero
}^{t=t_1}(1-\frac{\dot{S}(t)}{S^2(t)}\Sigma(\tzero , t))\kappa
\Phi(\overcirc{\gamma} (t))dt - \frac{S^3(\tzero )\dot{F}(\tzero
)}{2\Sigma (\tzero ,t_1)}\kappa\Phi(\overcirc{\gamma}(\tzero ))
\nonumber
\\ & \hspace*{-1.5em} \Bigl. - \int_{t=\tzero}^{t=t_1}\frac{\Sigma
(\tzero , t)}{2S(t)}\kappa \delta \rho (\overcirc{\gamma} (t))dt +
\frac{1}{\Sigma ( \tzero
,t_1)}\int_{t=\tzero}^{t=t_1}\frac{\Sigma^2(\tzero ,
t)}{2S(t)}\kappa \delta \rho (\overcirc{\gamma} (t))dt \Bigr\}
+O(\kappa^2) \label{deltamzMcV}\end{align} by successive
integration by parts and the use of (\ref{Gdef}), (\ref{Gdot}),
(\ref{Fdd}), (\ref{G}) and (\ref{FMcVIC}).  Again $\tzero$ is
given in terms of $z_1$ and $t_1$ by means of (\ref{red0}).
Equation (\ref{deltamzMcV}) is the correction to the apparent
magnitude redshift relation for Newtonian perturbations of FRW
cosmologies subject to the gauge fixing condition (\ref{FMcVIC}).

\section{The averaging procedure} \label{av}
\newcommand{\vzero}{\overcirc{v}}
\newcommand{\mylangle}{\raisebox{0ex}[1em][0.4em]{}\langle \hspace*{0.1ex} }
\newcommand{\myrangle}{\hspace*{0.5ex} \rangle \raisebox{0ex}[1em][0.4em]{}}
\newcommand{\Dt}{\raisebox{0ex}[1em][0ex]{\scriptsize $\! {\mathscr D}_t$}}
Suppose $\Phi$ is smooth on each level surface of $t$. Let
${\mathscr D}_t$ be a $3$-domain in one such surface. By
definition, the mean mass density perturbation on
 ${\mathscr D}_t$ is
\begin{equation} \mylangle \delta \rho \myrangle_{\Dt} :=
\frac{\int_{\Dt} \delta \hspace*{-.2ex} \rho \: d\vzero
\raisebox{0ex}[1em][0.8em]{} }{\int_{\Dt} d\vzero }
\label{deltarhoavV}\end{equation} where $d\vzero$ is the elemental
$3$-volume on ${\mathscr D}_t$ with respect to $\overcirc{\bf h}$.
By means of (\ref{deltarhoavV}) and (\ref{McVdeltarho}) one
obtains
\begin{equation} \mylangle  \delta \rho \myrangle_{\Dt} =
-\frac{3k}{S^3(t)}\mylangle\Phi\myrangle_{\Dt} -
\frac{\int_{\partial \Dt }
\prethree\nabla_{\overcirc{\mbox{\scriptsize \bf n}}} \Phi \:
d\vzero \raisebox{0ex}[1em][0.8em]{}}{S^3(t)\int_{\Dt }d\vzero }
+O(\kappa) \label{phiav1}
\end{equation} by means of the divergence theorem, where $\overcirc{\bf n}$
is the unit outward pointing normal to ${\mathscr D}_t$ at
$\partial {\mathscr D}_t$ with respect to $\overcirc{\bf h}$ in
the surface $\{t=\! \mbox{ \em const.}\}$, and where
\begin{equation} \mylangle \Phi \myrangle_{\Dt} := \frac{\int_{\Dt} \Phi \:
d\vzero \raisebox{0ex}[1em][0.8em]{}}{\int_{\Dt} d\vzero}
\end{equation} is the mean value of $\Phi$ on ${\mathscr D}_t$.
Note that $\mylangle \Phi \myrangle_{{\mathscr D}_t}$ is
independent of $t$. It will be assumed that the perturbed matter
distribution is sufficiently uniform on the large scale that, in
the limit ${\mathscr D}_t \rightarrow \infty$, the surface
integral term in (\ref{phiav1}) tends to zero, whilst $\mylangle
\delta \rho \myrangle_{{\mathscr D}_t}$ and $\mylangle \Phi
\myrangle_{{\mathscr D}_t}$ tend to limits $\mylangle \, \delta
\rho \myrangle_t$ and $\mylangle \Phi \myrangle$ respectively.
Under these conditions one obtains
\begin{equation}  \mylangle \, \delta \rho \myrangle_t =
-\frac{3k}{S^3(t)}\mylangle \Phi \myrangle +O(\kappa ) \ .
\label{phiav}\end{equation} In terms of the dimensionless density
parameter $\Omega$ of (\ref{Omegadef}) this gives
\begin{equation} \mylangle \delta \Omega \myrangle_t =
\frac{\kappa \mylangle \delta \rho
\myrangle_t}{3\overcirc{H}\Dummy^2(t)} + O(\kappa^2) =
(1+\overcirc{q}(t))\frac{\kappa\mylangle \Phi\myrangle
}{S(t)} + O(\kappa^2) \label{denparpert}\end{equation} by means of
(\ref{Hdot0}).  Equation (\ref{phiav}) and the right side of (\ref{denparpert})
must be considered invalid for $k=0$ since in that case $\mylangle \Phi
\myrangle$ must be infinite to give a finite $\mylangle \delta \rho
\myrangle_t$.

In order to obtain the mean correction to the apparent
magnitude-redshift relation, one replaces the function $\Phi
(\gamma (t))$ on the right of (\ref{deltamzMcV}) with its mean
value $\mylangle \Phi \myrangle$, and
$\left.\frac{d}{dt}\right|_{t=\tzero} \Phi (\gamma (t))$ by its
expectation value of zero.  In practice, since all cosmological
sources lie within gravitational potential wells, the
$\left.\frac{d}{dt}\right|_{t=\tzero} \Phi (\gamma (t))$ term in
(\ref{deltamzMcV}) will always give a positive contribution to the
apparent magnitude, but this will be so small as to be negligible
for present purposes.  By means of (\ref{deltarhoMcVt}),
(\ref{phiav}) and (\ref{denpardef}), and for $\tzero$ given in terms of $t_1$
and $z_1$
 by (\ref{red0}), one thus obtains
\begin{align} \mylangle \delta m(z_1)\myrangle & = \nonumber \\ &
\hspace*{-4em}-\frac{5}{2\ln 10} S^3(t_1)\overcirc{H}\Dummy^2 (t_1)
\mylangle \delta \Omega \myrangle_{t_1}
\left\{ -\frac{S^2(\tzero
)}{\dot{S}(t_1)\Sigma (\tzero ,
t_1)}\int_{t=\tzero}^{t=t_1}\dot{F}(t)dt +
3\int_{t=\tzero}^{t=t_1}\frac{\Sigma (\tzero , t)}{S^4(t)}dt
\right. \nonumber \\ & \hspace*{-0.5em}\left. -\frac{1}{\Sigma
(\tzero , t_1)}\int_{t=\tzero}^{t=t_1}\Sigma (\tzero ,
t)\dot{F}(t)dt - \frac{3}{\Sigma (\tzero ,
t_1)}\int_{t=\tzero}^{t=t_1}\frac{\Sigma^2(\tzero , t)}{S^4(t)}dt
\right\} +O(\kappa^2) \label{deltamzMcVav}\end{align} for the mean
correction to the apparent magnitude-redshift relation for
Newtonian perturbations of FRW cosmologies.  This is the
fundamental equation of the paper. It could alternatively have
been derived directly from (\ref{deltamz}) with the help of the
observation that the first term in the braces on the right of
(\ref{nuMcV}) gives a zero contribution to $\mylangle\delta m
(z_1)\myrangle$.  For nearby sources (\ref{deltamzMcVav}) gives
\begin{equation} \mylangle \delta m(z_1)\myrangle = -\frac{5}{4\ln
10}\mylangle \delta \Omega \myrangle_{t_1}z_1 +O(z_1^2) +
O(\kappa^2)  \label{deltamzMcVavapprox}\end{equation} by means of
(\ref{t1t0}) and (\ref{denparpert}).

In the particular case of a zero mean density perturbation,
$\mylangle \delta \rho \myrangle_t = 0$, equation
(\ref{deltamzMcVav}) clearly gives that there is a zero mean
correction to the apparent magnitude-redshift relation to first
order in $\kappa$.
 In the case of a non-zero mean density
perturbation $\mylangle \delta \rho \myrangle_t \neq 0$, one may
compare either with the background metric $\overcirc{\bf g}$ or
with that of the homogenised cosmology
determined from ${\bf g}$ by the constant potential
$\mylangle \Phi \myrangle$. (See \S\ref{ex}$\,a$.) In the latter
case the mean perturbation of the
potential is zero, so the mean density
perturbation is also zero.  Hence the correction to the apparent
magnitude-redshift
relation is zero, to first order in $\kappa$.

It is worthwhile to consider the case $\mylangle \delta \rho \myrangle_t >0$
in more detail.  Of the four terms in the braces on the right of
(\ref{deltamzMcVav}), the first three are positive.  And though the fourth
term is negative, it is evidently dominated by the second.  The sum of
the terms in the braces is therefore positive.  This shows that an object at
a given redshift appears brighter than in the reference FRW model.
Moreover an object at a given redshift in a high density universe appears
brighter than in a low density universe.  These effects can be regarded as
the gravitational lensing of the universe as a whole.

A fruitful approach to clumped matter perturbations of FRW
cosmologies, introduced by Dyer \& Roeder (1973), is to assume
that of all the matter present, a proportion $\alpha$ is
uniformly distributed and pressure-free, while the remaining
proportion $1-\alpha$ is gravitationally bound into clumps.  In
order to compute the lensing of light beams that remain far from
all clumps it suffices to take into consideration the effect of
only the uniformly distributed matter. In particular, shear can
be neglected.  The angular diameter distance (Schneider et al.\
1992, eq.\ (3.66)) determined from such clump-avoiding light
beams is often called the Dyer-Roeder distance.  For light beams
of large angular diameter, one expects the cumulative lensing
effect to approach that of the homogenised FRW cosmology.  The
transition between these two regimes has been studied by Linder
(1998) and found to occur typically between $1$ and $10$
arcseconds.

The Dyer-Roeder method may be implemented in the present framework as
follows. The background metric $\overcirc{\bf g}$ is taken to be the metric
of the homogenised FRW model.  The perturbation $\delta {\bf h}$ then
corresponds to the removal of a proportion $\alpha$ of the matter of this
model, followed by the addition of the same quantity of clumped matter.
The apparent magnitude-redshift relation for narrow, clump-avoiding light
beams is then given by (\ref{deltamzMcVav}), with $\langle \delta \Omega
\myrangle_{t_1}$ replaced by $-\alpha \mylangle \Omega \myrangle_{t_1}$.
By means of (\ref{denpardef}) this yields
\begin{align} \mylangle \delta m(z_1)\myrangle & = \nonumber \\ &
\hspace*{-4em} \frac{5}{2\ln 10} S^3(t_1)\overcirc{H}\Dummy^2 (t_1)
\alpha \mylangle \delta \Omega \myrangle_{t_1}
\left\{ -\frac{S^2(\tzero
)}{\dot{S}(t_1)\Sigma (\tzero ,
t_1)}\int_{t=\tzero}^{t=t_1}\dot{F}(t)dt +
3\int_{t=\tzero}^{t=t_1}\frac{\Sigma (\tzero , t)}{S^4(t)}dt
\right. \nonumber \\ & \hspace*{-0.5em}\left. -\frac{1}{\Sigma
(\tzero , t_1)}\int_{t=\tzero}^{t=t_1}\Sigma (\tzero ,
t)\dot{F}(t)dt - \frac{3}{\Sigma (\tzero ,
t_1)}\int_{t=\tzero}^{t=t_1}\frac{\Sigma^2(\tzero , t)}{S^4(t)}dt
\right\} +O(\kappa^2) \ . \label{angmz}\end{align}
Since the expression in the braces is positive for $z_1 >0$, this correction
is positive for $z_1 >0$.  Sources viewed at a given redshift along
clump-avoiding light beams thus appear dimmer than for an all-sky average.
This is as one would expect since clump-avoiding light beams are less
focussed.  For larger angular scales, the mean correction is no longer
given by (\ref{angmz}) and should be expected to tend to zero since the
mean density perturbation is zero.  Thus (\ref{angmz}) may be interpreted as
the mean correction to the apparent magnitude-redshift relation for narrow,
clump-avoiding light beams relative to that for wide angle beams.

The standard implementation of the Dyer-Roeder ansatz (e.g.\ Schneider et al.\
1992, p.138 et seq.) gives the Dyer-Roeder distance as a solution to the
Dyer-Roeder equation which describes light propagation through a uniformly
underdense region of space-time.  From the Dyer-Roeder distance one can
obtain an apparent magnitude-redshift relation.  On the other hand, an
apparent magnitude-redshift relation for light propagation through a uniformly
underdense region of space-time is also described by (\ref{angmz}) for an
appropriately valued constant $\mylangle \Phi \myrangle$
(see \ref{deltadenparun}).  Nonetheless one cannot expect these two relations
to agree unless similar gauge fixing conditions are applied in each case
(see (\ref{Gdotun}).

\section{Examples} \label{ex}
\begin{Example}[Uniform density perturbations]
In the special case \begin{equation} \Phi = \mbox{\em const.}
\label{phiun} \end{equation} the perturbation (\ref{McVdeltah})
reduces to \begin{equation} \delta h_{ij} = \kappa G(t) \Phi
\overcirc{h}_{ij} + O(\kappa^2) \ . \label{deltahun}
\end{equation}
To first order in $\kappa$ this is equivalent to
leaving the $3$-metric $\overcirc{\bf h}$ in (\ref{gFRW}) fixed
and perturbing the scale factor $S(t)$ according to
$S(t)\rightarrow (1+\frac{1}{2}G(t)\kappa\Phi )S(t)$.
The coordinate freedom (\ref{ttrans}), (\ref{xtrans}) reduces to the
freedom to change $t$ by an additive constant.
One sees directly from
(\ref{deltahun}) that the gauge condition (\ref{newgauge}) is satisfied iff
$G(t)$ satisfies  \begin{equation} \dot{G}(t_1) = 0  \ . \label{Gdotun}
\end{equation}

From (\ref{Tttphi}), (\ref{Titphi}) and
(\ref{Tijphi}) one has that the
perturbation is pressure-free to first
 order in $\kappa$, and so of a form as
discussed in {\S}\ref{Newt},
if $G(t)$ satisfies
\begin{equation} kG(t) =
\frac{1}{S(t)}(S^3(t)\dot{G}(t))\bracketdot + O(\kappa )
\label{GODEun} \ . \end{equation}
If $G(t)$ were to vanish at $t=t_1$ then, by (\ref{Gdotun}) and
(\ref{GODEun}), $G(t)$ would vanish for all $t$.  For consistency with
(\ref{Gfix}) one may choose
\begin{equation} G(t_1) = \frac{1}{S(t_1)} \ . \end{equation}

By (\ref{McVdeltarho}) and
(\ref{McVdeltap}) one has \begin{align} \delta \rho & =
-\frac{3k}{S^3(t)}\Phi + O(\kappa ) \label{deltarhoun} \\ \delta p
& = 0 + O(\kappa ) \ . \label{deltapun} \end{align}
By (\ref{denparpert}) the perturbation of the dimensionless
density parameter is given by \begin{equation} \delta\Omega (t) =
\frac{1}{2}(1+q(t))\frac{\kappa\Phi}{S(t)} +O(\kappa^2) \ .
\label{deltadenparun} \end{equation}
\end{Example}

\newcommand{\normalsizefrac}[2]{\frac{\mbox{\protect\normalsize
$#1$}}{\mbox{\protect\normalsize $#2$}}}
\begin{Example}[Point particle perturbations]
A perturbation by the
introduction of a family of comoving point particles with world
lines $\{{\bf x} = {\bf x}_{\alpha}\}$, $\alpha = 1,2,\ldots$ and
masses $m_{\alpha}>0$ is described by a density perturbation of
the form
\begin{equation} \delta \rho (t,{\bf x}) =
\frac{1}{S^3(t)}\sum_{\alpha}m_{\alpha}\prethree\overcirc{\delta}_{{\bf
x}_{\alpha}}({\bf x}) \label{ppdenpert} \end{equation} where
$\prethree\overcirc{\delta}_{{\bf x}_{\alpha}}({\bf x})$ is the
Dirac distribution with respect to $\overcirc{\bf h}$, centred on
${\bf x}_{\alpha}$, on each level surface of $t$. The particles
shall represent the galaxies.

In order to satisfy (\ref{ppdenpert}) and (\ref{McVdeltarho}) one
seeks a potential $\Phi ({\bf x})$ satisfying \begin{equation}
-(\prethree\overcirc{\Delta}\Phi ({\bf x}) + 3k\Phi ({\bf x}) ) =
\sum_{\alpha}m_{\alpha}\prethree\overcirc{\delta}_{{\bf
x}_{\alpha}}({\bf x}) \ . \label{ppPDE} \end{equation}  Since this
equation is linear one may decompose $\Phi({\bf x})$ as a sum
\begin{equation} \Phi ({\bf x}) = \sum_{\alpha}\Phi_{\alpha}({\bf
x}) \ . \label{phisum} \end{equation}  In order to consider a
typical summand $\Phi_{\alpha}({\bf x})$ it is convenient to
express the $3$-metric $\overcirc{\bf h}$ in the form
\begin{equation} \overcirc{\bf h} = \begin{cases}
\frac{1}{k}(d\omega_{\alpha}^2 +
\sin^2\omega_{\alpha}d\Omega_{\alpha}^2) & \text{if $k>0$}\\[1ex]
dr_{\alpha}^2 + r_{\alpha}^2d\Omega_{\alpha}^2 & \text{if
$k=0$}\\[1ex] \frac{1}{(-k)}(d\omega_{\alpha}^2 +
\sinh^2\omega_{\alpha}d\Omega_{\alpha}^2) & \text{if $k<0$}
\end{cases} \end{equation} where $d\Omega_{\alpha}^2$ is the
$2$-sphere metric and the radial coordinate $\omega_{\alpha}$ is
defined by \begin{equation} \omega_{\alpha}(r_{\alpha}) := \begin{cases}
2\tan^{-1}\left(\frac{\sqrt{k}}{2}r_{\alpha}\right) & \text{if $k>0$}
\\[1ex] 2\tanh^{-1}\left( \frac{\sqrt{-k}}{2}r_{\alpha}\right) & \text{if
$k<0$} \end{cases} \end{equation} with the origin
$\omega_{\alpha}=0$ being the point ${\bf x}_{\alpha}$ of
(\ref{ppdenpert}).  The range of $\omega_{\alpha}$ is $0\leq
\omega_{\alpha} \leq \pi$ if $k>0$ and $0\leq \omega_{\alpha}
<\infty$ if $k < 0$.

The general radial solution to (\ref{ppPDE}), as found by Newman
\& McVittie (1982), is \begin{equation} \Phi_{\alpha} ({\bf x}) =
\begin{cases} \normalsizefrac{\sqrt{k}}{4\pi \sin \omega_{\alpha}} (
m_{\alpha} \cos 2\omega_{\alpha} + C_{\alpha}\sin 2\omega_{\alpha}
) & \text{if $k>0$}\\[2ex] \normalsizefrac{m_{\alpha}}{ 4\pi r} +
C_{\alpha} & \text{if $k=0$}\\[1.5ex]
\normalsizefrac{\sqrt{-k}}{4\pi \sinh \omega_{\alpha}}(m_{\alpha}
\cosh 2\omega_{\alpha} + C_{\alpha}\sinh 2\omega_{\alpha} ) &
\text{if $k<0$}
\end{cases} \label{phirad} \end{equation} where $C_{\alpha}$ is an arbitrary
constant.

The case $k>0$ would appear to be the simplest insofar as the
level surfaces of $t$ are compact, so physical plausibility
demands that there are at most finitely many particles present.
However it is evident from (\ref{phirad}) in this case that for
each particle there is a complementary particle of equal mass
located at the antipodal point in each surface $\{t=\mbox{\em
const.}\}$.  This bizarre doppelg{\"a}nger phenomenon leads one to
question whether the $k>0$ solution is realistic after all.  It is
unclear whether the problem is an artifact of the symmetry of the
level surfaces of $t$ or of the linear approximation.  There may
even be a deeper issue here concerning the constraint components
of the Einstein equations (D'Eath 1976).

For $k>0$ an integration of (\ref{ppPDE}) over a level surface of
$t$ with respect to the volume element $d\overcirc{v}$ associated
with $\overcirc{\bf h}$ yields \begin{equation}
-3k\int_{t=\mbox{\scriptsize \em const.}}\Phi \, d\overcirc{v} =
\sum_{\alpha}m_{\alpha} \ .\end{equation} Setting
$\prethree\overcirc{V}:= \int_{t=\mbox{\scriptsize \em const.}}
d\overcirc{v}$ one thus obtains
\begin{equation} \mylangle \delta \rho \myrangle_t :=
\frac{\sum_{\alpha}m_{\alpha}}{S^3(t)\prethree \overcirc{V}} =
-\frac{3k}{S^3(t)}\mylangle \Phi \myrangle \label{phiavppkpos}
\end{equation}
for
\begin{equation} \mylangle \Phi \myrangle =
\frac{\int_{t=\mbox{\scriptsize \em const.}}\Phi \,
d\overcirc{v}}{\int_{t=\mbox{\scriptsize \em
const.}}d\overcirc{v}} \ . \end{equation}
Note that (\ref{phiavppkpos}) agrees with (\ref{phiav}) even though $\Phi$
is not smooth in the present case.

In the case $k<0$ case the surfaces $\{t=\mbox{\em const.}\}$ have
infinite volume, so infinitely many particles (or none) are needed
in order to achieve a distribution that is uniform on the large
scale.  For each $\alpha$ one must choose $C_{\alpha}=
-m_{\alpha}$ in order that $\Phi_{\alpha}$ decays to zero at
infinity. Each $\Phi_{\alpha} ({\bf x})$ is then integrable on the
level surfaces $\{ t=\mbox{\em const.}\}$ and a simple reciprocity
argument indicates that for a sufficiently uniform distribution of
particles $\Phi ({\bf x}) = \sum_{\alpha} \Phi_{\alpha} ({\bf x})$
should converge everywhere other than on the world lines of the
particles.  Let ${\mathscr D}_t$ be a compact $3$-domain in a
level surface of $t$. An integration of (\ref{ppPDE}) over
${\mathscr D}_t$ with respect to $d\overcirc{v}$ yields
\begin{equation} -3k\int_{\Dt} \Phi \, d\overcirc{v} =
\sum_{\alpha} m_{\alpha} +\int_{\partial \Dt }
\overcirc{\nabla}_{\smashovercirc{\bf n}}\Phi \, d\overcirc{a}
\end{equation} where $\overcirc{\bf n}$ is the outward unit normal
at $\partial {\mathscr D}_t$, $d\overcirc{a}$ is the area element
on $\partial {\mathscr D}_t$ induced by $\overcirc{\bf h}$, and
where, in the first term on the right, the sum is carried over all
$\alpha$ such that ${\bf x}_{\alpha} \in {\mathscr D}_t$.  Setting
$\overcirc{V}({\mathscr D}_t) := \int_{{\mathscr D}_t}
d\overcirc{v}$ one obtains
\begin{equation}
\mylangle \delta \rho \myrangle_{\Dt } =
-\frac{3k}{S^3(t)}\mylangle \Phi \myrangle_{{\mathscr D}_t}
-\frac{\int_{\partial \Dt } \overcirc{\nabla}_{\overcirc{\bf
n}}\Phi \, d\overcirc{a}}{S^3(t)\prethree\overcirc{V}({\mathscr D}_t)}
\label{ppdenpertkneg}
\end{equation}
where
\begin{equation} \mylangle \delta \rho \myrangle_{{\mathscr D}_t} :=
\frac{\sum_{\alpha}m_{\alpha}}{S^3(t)\prethree\overcirc{V}({\mathscr
D}_t)}
\end{equation}
is the mean density perturbation on ${\mathscr D}_t$ and
\begin{equation} \mylangle \Phi \myrangle_{{\mathscr D}_t} := \frac
{\int_{{\mathscr D}_t}\Phi \,
d\overcirc{v}}{\prethree\overcirc{V}({\mathscr D}_t)}
\end{equation} is the mean value of $\Phi$ on ${\mathscr D}_t$.
Reasonable uniformity conditions on the distribution of the
particles should ensure that $\mylangle \delta \rho
\myrangle_{{\mathscr D}_t}$  and $\mylangle \Phi
\myrangle_{{\mathscr D}_t}$ tend to limits $\mylangle \delta \rho
\myrangle_t$ and $\mylangle \Phi \myrangle$ respectively for
arbitrarily large ${\mathscr D}_t$.  Such conditions should also
ensure that the second term on the right of (\ref{ppdenpertkneg})
tends to zero for large ${\mathscr D}_t$.  One then obtains
(\ref{phiav}) for $k<0$, again even though $\Phi$ is not smooth.

In the $k=0$ case $\Phi_{\alpha}(r_{\alpha})$, as given by
(\ref{phirad}), decays to zero at infinity only if $C_{\alpha}=0$,
and then only as $1/r_{\alpha}$.  For a uniform distribution of
particles the potential $\Phi = \sum_{\alpha}\Phi_{\alpha}$ would
be infinite everywhere, so the theory breaks down in this case.

For $k\neq 0$ the perturbation of the space-time metric
corresponding to the introduction of the uniform distribution of
comoving point particles is described by (\ref{gdef}),
(\ref{McVdeltah}), (\ref{phisum}) and (\ref{phirad}), and the mean
correction to the apparent magnitude-redshift relation, relative
to the background metric, is given by (\ref{deltamzMcVav}).
\end{Example}

\begin{Example}[Swiss cheese model]
This model, proposed by Einstein \& Strauss (1945,1946), is an
exact $C^{1-}$ solution to the Einstein equations consisting of a
pressure-free FRW model in which spherical regions are replaced by
spherical pieces of Schwarzschild geometry. The idea is that each
of these spherical `holes' represents the condensation of dust
into a star, represented by the singularity at the centre. For
present purposes, the singularities will be considered to
represent the galaxies and the intervening dust, the `cheese',
will represent the intergalactic medium. The model will be
considered here in terms of the perturbative formalism of \S
\ref{Newt}.

Let $\rho (t)$ be the density of the intergalactic dust.  In order
to describe a spherically symmetric hole with a central point
particle of mass $m$ and coordinate radius $\omega = \hat{\omega}$
if $k\neq0$, or $r=\hat{r}$ if $k=0$, one may seek a radial potential
function $\Phi ({\bf x})$ which satisfies
\begin{equation} -(\prethree\Delta \Phi ({\bf x}) + 3k\Phi ({\bf x})) =
-S^3(t_1)\rho (t_1) + m\prethree\delta_0({\bf x})
\label{PDEhole}\end{equation} in the hole and matches in a
$C^{2-}$ manner to a constant potential $\Phi ({\bf x}) =
\hat{\Phi}$ outside the hole.  The first term on the right of
(\ref{PDEhole}) ensures that the hole is a vacuum. The general
radial solution to (\ref{PDEhole}) has the form \begin{equation}
\Phi ({\bf x}) =
\begin{cases} \normalsizefrac{\sqrt{k}m}{4\pi}\normalsizefrac{\cos
2\omega}{\sin \omega} +
\normalsizefrac{C}{8\pi}\normalsizefrac{\sin 2\omega}{\sin\omega}
+ \normalsizefrac{S^3(t_1)\rho(t_1)}{3k} & \text{if $k>0$} \\[2ex]
\normalsizefrac{m}{4\pi r} + \normalsizefrac{C}{4\pi} +
\normalsizefrac{1}{6}S^3(t_1)\rho(t_1)r^2 & \text{if $k=0$}
\\[2ex] \normalsizefrac{\sqrt{-k}m}{4\pi}\normalsizefrac{\cosh
2\omega}{\sinh 2\omega}
+\normalsizefrac{C}{8\pi}\normalsizefrac{\sinh 2\omega}{
\sinh\omega} + \normalsizefrac{S^3(t_1)\rho (t_1)}{3k} & \text{if
$k<0$} \end{cases} \label{phihole}\end{equation} within the hole,
where $C$ is a constant.

In order for the solution (\ref{phihole}) to join in a $C^{2-}$
manner to the constant solution $\Phi ({\bf x}) = \hat{\Phi}$
outside the hole, the radial derivative of $\Phi ({\bf x})$ must
vanish at the boundary. By means of the divergence theorem, an
integration of (\ref{PDEhole}) over the hole therefore yields
\begin{equation} m = S^3(t_1)\rho
(t_1)\prethree\overcirc{V}_{\mbox{\scriptsize \em hole}} -
3k\int_{\mbox{\scriptsize \em hole}}\Phi\, d\overcirc{v}
\label{mhole}\end{equation} for all values of $k$, where
$\prethree\overcirc{V}_{\mbox{\scriptsize \em
hole}}:=\int_{\mbox{\scriptsize \em hole}}d\overcirc{v}$ is the
volume of the hole with respect to $\overcirc{\bf h}$. Thus the
mean matter density of the hole is precisely $\rho (t_1)$ for
$k=0$ and $\rho (t_1) + O(\hat{\omega}\dummy^2)$ for $k\neq 0$.
Substitution of (\ref{phihole}) into (\ref{mhole}) yields
\begin{equation} C= \begin{cases}
-2\sqrt{k}m\normalsizefrac{(\cos\hat{\omega} - \frac{1}{3}\cos
3\hat{\omega})}{(\sin\hat{\omega} - \frac{1}{3}\sin3\hat{\omega})}
& \text{if $k>0$}
\\[2ex]
-2\sqrt{-k}m \normalsizefrac{ (\frac{1}{3}\cosh 3\hat{\omega} -
\cosh\hat{\omega})}{(\frac{1}{3}\sinh 3\hat{\omega} -
\sinh\hat{\omega})} & \text{if $k<0$} \ ,
\end{cases} \label{Chole}
\end{equation} whilst in the case $k=0$ (\ref{mhole}) gives
\begin{equation} m =
\frac{4\pi}{3}\rho (t_1)S^3(t_1)\hat{r}\dummy^3 \quad \mbox{ if
$k=0$\ .} \label{mhole0}\end{equation}

For $k\neq 0$ the continuity of $\Phi$ at the boundary of the hole
gives \begin{equation} \hat{\Phi} = \begin{cases}
\normalsizefrac{S^3(t_1)\rho(t_1)}{3k} -
\normalsizefrac{\sqrt{k}m}{3\pi}\normalsizefrac{1}{(\sin
\hat{\omega} - \frac{1}{3}\sin 3\hat{\omega})} &\text{if $k>0$}
\\[2ex] \normalsizefrac{S^3(t_1)\rho (t_1)}{3k} +
\normalsizefrac{\sqrt{-k}m}{3\pi}\normalsizefrac{1}{(\frac{1}{3}\sinh
3\hat{\omega} - \sinh \hat{\omega})} & \text{if $k<0$} \end{cases}
\label{phihat} \end{equation} by means of (\ref{phihole}) and
(\ref{Chole}), whilst in the case $k=0$ one obtains
\begin{equation} \hat{\Phi} = \frac{3m}{8\pi \hat{r}} +
\frac{C}{4\pi}\quad \text{if $k=0$} \label{phihat0} \end{equation}
by means of (\ref{phihole}) and (\ref{mhole}).

Suppose now that there are many holes, each labelled by an index
$\alpha$.  Let ${\mathscr D}_t$ be a compact $3$-domain in a level
surface of $t$ such that $\partial {\mathscr D}_t$ intersects none
of the holes.  The mean value of $\Phi$ on ${\mathscr D}_t$ is
\begin{equation} \mylangle\Phi\myrangle_{{\mathscr D}_t} :=
\frac{\int_{{\mathscr D}_t}\Phi\, d\overcirc{v}}{\prethree
\overcirc{V}({\mathscr D}_t)} = \hat{\Phi} +
\frac{\sum_{\alpha}\int_{\mbox{\scriptsize \em
hole}_{\alpha}}(\Phi - \hat{\Phi} )\, d\overcirc{v}}{\prethree
\overcirc{V}({\mathscr D}_t)}
\end{equation} where $\prethree \overcirc{V}({\mathscr D}_t) :=
\int_{{\mathscr D}_t} d\overcirc{v}$ is the volume of ${\mathscr
D}_t$ with respect to $\overcirc{\bf h}$ and the sum is carried
out over all $\alpha$ such that the $\alpha^{\mbox{\scriptsize \em
th}}$ hole is contained in ${\mathscr D}_t$. By means of
(\ref{mhole}) and (\ref{phihat}) in the cases $k\neq 0$, and by
means of (\ref{phihole}), (\ref{mhole0}) and (\ref{phihat0}) in
the case $k=0$, one has
\begin{equation} \mylangle \Phi \myrangle_{{\mathscr D}_t} = \hat{\Phi} +
M_{\smash[b]{{\mathscr D}_t}} \end{equation} where
\begin{equation} M_{\smash[b]{{\mathscr D}_t}} =\begin{cases}
\normalsizefrac{1}{\prethree V({\mathscr D}_t)}{\cdot}
\frac{1}{3}\sum_{\alpha}\normalsizefrac{m_{\alpha}}{k}\{
\normalsizefrac{2(\hat{\omega}_{\alpha}-\frac{1}{2}\sin
2\hat{\omega}_{\alpha})}{(\sin\hat{\omega}_{\alpha}-\frac{1}{3}\sin
3\hat{\omega}_{\alpha})} -1 \} & \text{if $k>0$}
\\[2ex]\normalsizefrac{1}{\prethree V({\mathscr D}_t)}{\cdot}
\frac{1}{10}\sum_{\alpha}m_{\alpha}\hat{r}_{\alpha}\dummy^2
&\text{if $k=0$} \\[2ex]  \normalsizefrac{1}{\prethree V({\mathscr
D}_t)} {\cdot}\frac{1}{3}\sum_{\alpha}\normalsizefrac{m_{\alpha}}{k}\{ 1
- \normalsizefrac{2(\frac{1}{2}\sinh 2\hat{\omega}_{\alpha} -
\hat{\omega}_{\alpha})}{(\frac{1}{3}\sinh 3 \hat{\omega}_{\alpha}
- \sinh \hat{\omega}_{\alpha})}\} & \text{if $k<0$ .}
\end{cases} \end{equation}  It will be assumed that the
distribution of holes is sufficiently uniform that
$M_{\smash[b]{{\mathscr D}_t}}$ and $\mylangle \Phi
\myrangle_{{\mathscr D}_t}$ tend to limits $M$ and $\mylangle \Phi
\myrangle$ respectively as ${\mathscr D}_t$ becomes arbitrarily
large. In order to obtain $\mylangle \Phi \myrangle = 0$ one must
have
\begin{equation} \hat{\Phi} +M =0 \ .
\end{equation}  By (\ref{phiav}) one then has \begin{equation}
\mylangle \delta \rho \myrangle_t = 0 + O(\kappa ) \ .
\end{equation}

In the case $k=0$ the quantities $\hat{m}_{\alpha}$ and
$\hat{r}_{\alpha}$ are, for each $\alpha$, related by an equation
of the form (\ref{mhole}), whereby one has \begin{equation}
\frac{M_{{\mathscr D}_t}}{S(t_1)} = \left(\frac{3\pi}{4\rho
(t_1)}\right)^{2/3}
\frac{\sum_{\alpha}m^{5/3}_{\alpha}}{S^3(t_1)\prethree
\overcirc{V}({\mathscr D}_t)} \quad  \quad \mbox{if $k=0$} \ .
\label{Momega}\end{equation}

For $k\neq 0$ the quantities $\hat{m}_{\alpha}$ and
$\hat{r}_{\alpha}$ are, for each $\alpha$, related by an equation
of the form (\ref{phihat}) which involves the constant
$\hat{\Phi}$.  However the holes may be assumed sufficiently small
that (\ref{Momega}) is a valid approximation for $k\neq 0$.  By
(\ref{deltadenparun}) and (\ref{Momega}), the dimensionless
density parameter of the intergalactic matter is then to be
perturbed by an amount
\begin{equation} \delta
\Omega_{\mbox{\scriptsize \em cheese}} =
-(1+q(t_1))\frac{\kappa M}{S(t_1)} \quad \quad\mbox{if
$k\neq 0$} \label{sigmacheese}
\end{equation}  for
\begin{equation} \frac{M}{S(t_1)} = \left( \frac{3\pi}{4\rho
(t_1)}\right)^{2/3}\lim_{{\mathscr D}_{t_1}\rightarrow
\infty}\frac{\sum_{\alpha}m_{\alpha}^{5/3}}{\overcirc{V}_{t_1}({\mathscr
D}_{t_1})} \ , \label{Minfty} \end{equation} the sum being carried
over all $\alpha$ for which the $\alpha^{\mbox{\scriptsize\em
th}}$ hole is contained in ${\mathscr D}_{t_1}$, with
$\overcirc{V}_{t_1}({\mathscr D}_{t_1}) =
S^3(t_1)\prethree\overcirc{V}(\smash[b]{{\mathscr D}_{t_1}})$ the
volume of ${\mathscr D}_{t_1}$ with respect to the $3$-metric
$S^3(t_1)\overcirc{\bf h}$ induced by $\overcirc{\bf g}$ on
$\{t=t_1\}$. If one regards the $k=0$ case as a limit of the
$k\neq 0$ cases then (\ref{sigmacheese}) may considered to apply
for all $k$.

It is the perturbation (\ref{sigmacheese}) of the density of the
intergalactic matter that distinguishes the present approach to
the Swiss cheese model from those of other authors, including Dyer
\& Roeder (1974). Only with such a perturbation will the mean
density parameter $\mylangle \Omega \myrangle_t$ be unperturbed.
And only then, as discussed in {\S}\ref{Newt}, will there be a zero
mean correction to the apparent magnitude-redshift relation to
first order in $\kappa$.
\end{Example}

\section{Concluding remarks} \label{conc}
The validity of the linear approximation employed in this paper
depends upon the smallness of the mean dimensionless density
perturbation $\mylangle \delta \Omega \myrangle_t$ associated with
the matter in the galaxies.  Astronomical estimates of this
quantity are at present inconclusive, although there is a general
consensus that one does have $\mylangle \delta \Omega \myrangle_t
\ll 1$. If this is correct then the theory presented should
provide a valid description of the total gravitational field and
cumulative weak lensing effects of the galaxies for any given
model of the distribution of galactic matter. Depending upon the
actual value of $\mylangle \delta \Omega \myrangle_t$ and the
range of redshifts of interest, there may be a need to carry the
theory to second or higher order in $\kappa$. This would at least
take into account the effect of shear on the apparent brightness
of cosmological sources.  Caustics however cannot be adequately
described by any finite order power series analysis. For this it
would be necessary to consider the full non-linear theory.

\vspace{3ex} {\small \noindent I express my appreciation
to Prof.\ M.A.H. MacCallum for his interest and encouragement,
and to the Department of Mathematics and Applied Mathematics at the
University of Cape Town for their hospitality.
Funding was provided by the Swedish Natural Science Research
Council (NFR).}

\label{lastpage}

\end{document}